\documentclass[fleqn,usenatbib]{mnras}

\usepackage[T1]{fontenc}
\usepackage{ae,aecompl}

\usepackage{graphicx,color}	
\usepackage{amsmath}	
\usepackage{amssymb}	
\usepackage{soul}

\newcommand\mearth{{\,{\rm M}_{\oplus}}}
\newcommand\mj{{\,{\rm M}_{\rm J}}}

\def\jh#1{\textcolor{magenta}{#1}}

\title[Making sub-Jovian planets via GI]{On the origin of wide-orbit ALMA planets: giant protoplanets disrupted by their cores}

\author[]{J. Humphries\thanks{rjh73@le.ac.uk}  \& S. Nayakshin
\\
\\
Department of Physics and Astronomy, University of Leicester, Leicester LE1 7RH, UK.
}

\date{Accepted XXX. Received YYY; in original form ZZZ}

\pubyear{2018}

\begin{document}
\label{firstpage}
\pagerange{\pageref{firstpage}--\pageref{lastpage}}
\maketitle

\begin{abstract}
Recent ALMA observations may indicate a surprising abundance of sub-Jovian planets on very wide orbits in protoplanetary discs that are only a few million years old.  These planets are too young and distant to have been formed via the Core Accretion (CA) scenario, and are much less massive than the gas clumps born in the classical Gravitational Instability (GI) theory. It was recently suggested that such planets may form by the partial destruction of GI protoplanets: energy output due to the growth of a massive core may unbind all or most of the surrounding pre-collapse protoplanet.
Here we present the first 3D global disc simulations that simultaneously resolve grain dynamics in the disc and within the protoplanet. 
We confirm that massive GI protoplanets may self-destruct at arbitrarily large separations from the host star provided that solid cores of mass $\sim $10-20 $ M_{\oplus}$ are able to grow inside them during their pre-collapse phase. 
In addition, we find that the heating force recently analysed by \cite{MassetVelascoRomero17} perturbs these cores away from the centre of their gaseous protoplanets.  This leads to very complicated dust dynamics in the protoplanet centre, potentially resulting in the formation of multiple cores, planetary satellites, and other debris such as planetesimals within the same protoplanet. A unique prediction of this planet formation scenario is the presence of sub-Jovian planets at wide orbits in Class 0/I protoplanetary discs.

\end{abstract}


\section{Introduction}

It is now widely believed that the gaps in the $\sim $ 1 mm dust discs observed by the Atacama Large Millimeter Array (ALMA) are the signatures of young planets \citep{Alma2015,IsellaEtal16,LongEtal18,DSHARP1,DSHARP7}. Modelling suggests that these planets are often wide-orbit Saturn  analogues \citep[][]{DipierroEtal16a,ClarkeEtal18,LodatoEtal19}. Such planets present a challenge for both of the primary planet formation theories, albeit for different reasons. 

In the classical planetesimal-based Core Accretion (CA) theory \citep{PollackEtal96,IdaLin04a} forming massive solid cores at wide separations in a $\sim 1$~Myr old disc is challenging as the process is expected to take more than an order of magnitude longer than this \citep[e.g.,][]{KL99}. Core growth via pebble accretion is much faster; however the process is not very efficient \citep{OrmelLiu18, LinEtal18} and may require more pebbles than the observations indicate. Additionally, in any flavor of CA scenario the ALMA planets should be in the runaway gas accretion phase. This process is expected to produce planets much more massive than Jupiter {\em very rapidly} at these wide orbits, which does not seem to be the case \citep{NayakshinEtal19}. \cite{NduguEtal19} has very recently detailed these constraints. As much as $2000 \mearth$ of pebbles in the disc are required to match the ALMA gap structures, and the resulting planet mass function indeed shows too few sub-Jovian planets and too many $M_{\rm p} > 1 \mj$ planets.

In the Gravitational Instability (GI) scenario \citep{Boss98, Rice05,Rafikov05} planets form very rapidly, e.g., in the first $\sim 0.1$ Myr \citep{Boley09}. The age of ALMA planets is thus not a challenge for this scenario; ALMA planets if anything, are `old' for GI. However, the minimum initial mass of protoplantary fragments formed by GI in the disc is thought to be at least $\sim 1 \mj$ \citep{BoleyEtal10}, and perhaps even $\sim 3-10 \mj$ \citep{KratterEtal10,ForganRice13b,KratterL16}. This is an order of magnitude larger than the typical masses inferred for the ALMA gap opening planets \citep{NayakshinEtal19}.  

Formation of a protoplanetary clump in a massive gas disc is only the first step in the life of a GI-made planet, and its eventual fate depends on many physical processes \citep[e.g., see the review by][]{Nayakshin_Review}. In this paper we extend our earlier work \citep{HumphriesNayakshin18} on the evolution of GI protoplanets in their pebble-rich parent discs.
Newly born GI protoplanets are initially very extended with radii of $\sim$ 1 AU. After a characteristic cooling time, Hydrogen molecules in the protoplanet dissociate and it collapses to a tightly bound Jupiter analogue \citep{Bodenheimer74,HelledEtalPP62014}. During this cooling phase protoplanets are vulnerable to tidal disruption via interactions with the central star or with other protoplanets, which may destroy many of these nascent planets if they migrate to separations closer than 20 AU \citep{HumphriesEtal19}. 

Additionally, pebble accretion plays a key role in the evolution of GI protoplanets. Observations \citep{TychoniecEtal18} suggest that young Class 0 protoplanetary discs contain as much as hundreds of Earth masses in pebble-sized grains. These will be focused inside protoplanets as they are born \citep{BoleyDurisen10} and accreted during any subsequent migration \citep{BaruteauEtal11,JohansenLacerda10,OrmelKlahr10}. Both analytical work \citep{Nayakshin15a} and 3D simulations \citep{HumphriesNayakshin18} demonstrate that protoplanets accrete the majority of mm and above sized dust grains that enter their Hill spheres, considerably enhancing their total metal content. 
Once accreted, these pebbles are expected to grow and settle rapidly inside the protoplanet, likely becoming locked into a massive core \citep{Kuiper51b,BossEtal02,HS08,BoleyEtal10,Nayakshin10a,Nayakshin10b}. 

\cite{Nayakshin16a} showed via 1D population synthesis models that core formation inside young GI protoplanets may in fact release enough heat to remove some or all of their gaseous envelopes. This process could therefore downsize GI protoplanets, making GI a more physically plausible scenario for hatching ALMA planets. This idea is also attractive because of parallels with other astrophysical systems, e.g. galaxies losing much of their initial gaseous mass due to energetic feedback from supermassive black holes \citep{DiMatteo05}.

In \cite{HumphriesNayakshin18} we used a sink particle approximation to study the accretion of pebbles onto GI protoplanets. The sink particle approach provides a reliable estimate for the total mass of pebbles captured by a protoplanet, but it does not allow us to study their subsequent evolution. 
In this paper we improve on our previous work by modelling the protoplanet hydrodynamically, thus resolving  both gas and dust dynamics within it. Nevertheless, it remains beyond current computational means to resolve the growing solid core in such simulations, and so we introduce a {\em dust-sink} in the protoplanet centre to model the core. As pebbles accrete onto the core, we pass the liberated gravitational potential energy to the surrounding gas in the protoplanet. Our simulations therefore also aim to explore the effects of core feedback as introduced in \cite{Nayakshin16a}.

The paper is structured as follows. In Section \ref{sec:analytic_fb} we briefly calculate the expected core mass necessary to disrupt a young GI protoplanet. Following this, in Section \ref{sec:methods} we describe the new physics added to our previous simulations \citep{HumphriesNayakshin18} in order to numerically model core growth and feedback. We also study the settling timescale for a variety of grain sizes inside protoplanets. 
In Section \ref{sec:results} we present the main results of the paper, examining how feedback driven disruption over a range of feedback timescales and pebble to gas ratios can unbind protoplanets and leave behind rocky cores at tens of AU. In Section \ref{sec:core} we extend this analysis and take a closer look at the core during the feedback process. In Section \ref{sec:discussion} we outline the observational implications of protoplanet disruption process and also discuss some of the limitations of our model. Finally, we summarise the conclusions of our paper in Section \ref{sec:conclusions}: if rocky cores rapidly form inside GI protoplanets, the resultant release of energy may disrupt these objects and leave super-Earth and potentially Saturn mass cores stranded at tens to hundreds of AU.

\section{Analytical Estimates}
\label{sec:analytic_fb}
\begin{figure}
\includegraphics[width=0.99\columnwidth]{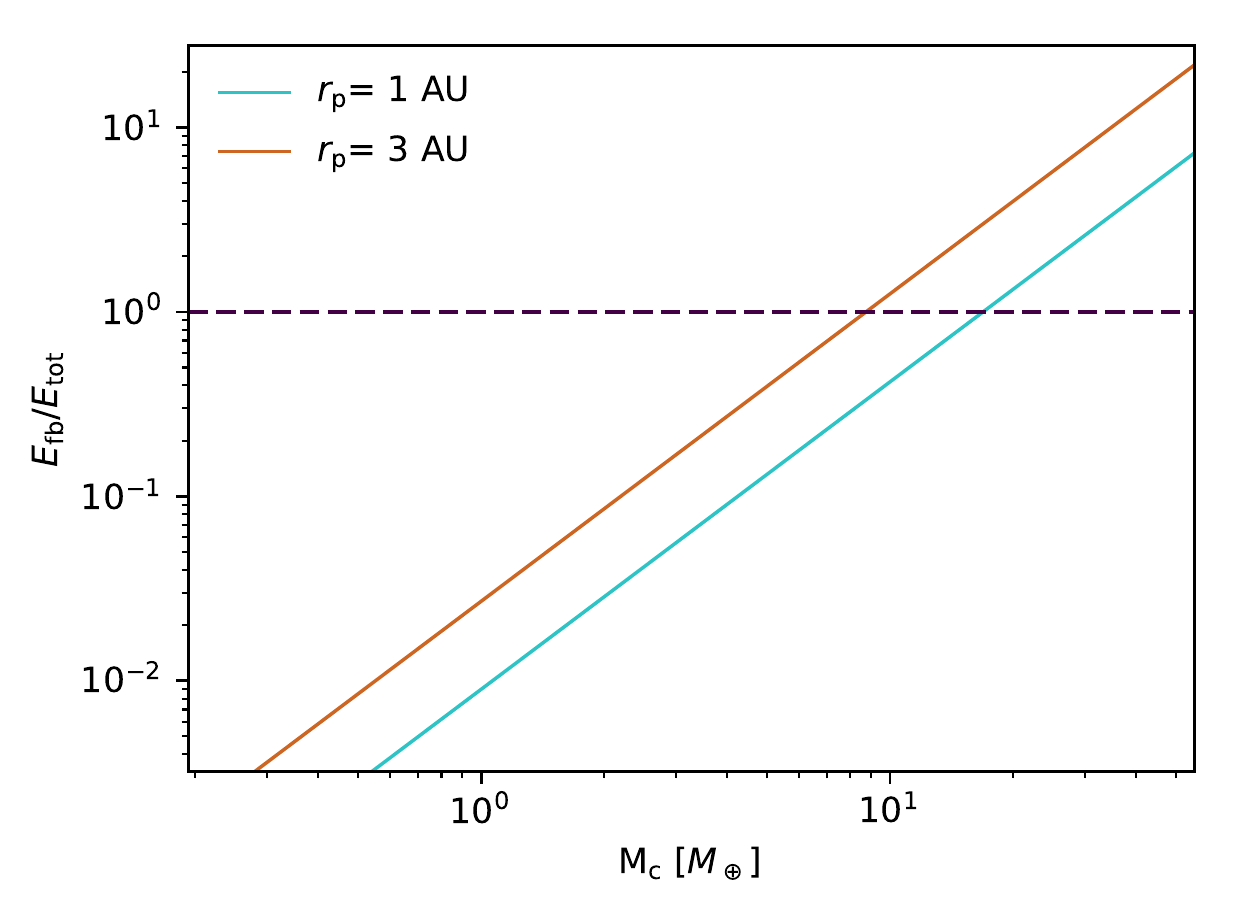}
\caption{Liberated gravitational potential energy from core formation as a fraction of protoplanet binding energy from Equation \ref{eq:feedback}. The protoplanet has a mass of 5$M_J$ and a solid core central density of 5 gcm$^{-3}$. It takes a core of 17$M_{\oplus}$ to unbind this protoplanet if it has a radius of 1 AU.}
\label{fig:an_feedback}
\end{figure}

Here we estimate the core mass needed to unbind a newly born GI protoplanet with a radius of a few AU, which is typical for pre-collapse planets. We assume that the protoplanet can be modelled as a polytropic sphere with mass $M_{\rm p}$, radius $r_{\rm p}$, and adiabatic index $\gamma=7/5$. The Virial theorem tells us that the total energy of a polytrope ($E_{\rm tot}$) is half of its gravitational binding energy and is therefore given by

\begin{equation}
E_{\rm tot} = -\dfrac{3GM_{\rm p}^2}{5r_{\rm p}}.
\label{eq:binding}
\end{equation}

After its birth, this protoplanet is able to accrete gas and dust from the disc. If accreted dust is able to grow large enough, it will rapidly settle to the centre and form a core \citep{HelledEtal08,Nayakshin10a,Nayakshin10b}. This process will release gravitational potential energy and heat the central regions of the protoplanet. Assuming that the accretion energy of the core is released into the surrounding gas instantaneously, as is done in the Core Accretion scenario \citep[e.g.,][]{Stevenson82,PollackEtal96,MordasiniEtal12}, the total corresponding feedback energy ($E_{fb}$) increases at the rate 

\begin{equation}
E_{\rm fb} = \int_0^t \dfrac{G M_{\rm c} \dot M_{\rm c}}{r_{\rm c}} dt',
\label{eq:E_fb}
\end{equation}

where $M_{\rm c}$ and $r_{\rm c}$ are the mass and radius of the luminous core. If we assume that the rate of dust accretion onto the core is constant and we take a constant core density $\rho_{\rm c}$, we can rewrite $r_{\rm c}$ in terms of $M_{\rm c}$ and integrate over $t$ to find $E_{\rm fb}$ as 

\begin{equation}
E_{\rm fb} =  \dfrac{3}{5}  \left( \dfrac{4 \pi \rho_{\rm c} }{3} \right)^{1/3} G (M_{\rm c})^{5/3},
\end{equation}

where we have assumed that all of the gravitational potential energy is liberated instantaneously. We can then study this feedback energy as a fraction of the total energy of the protoplanet from Equation \ref{eq:binding},

\begin{equation}
\dfrac{E_{\rm fb}}{E_{\rm tot}} 
= 1.3 \left(\dfrac{\rho_{\rm c}}{5 \rm gcm^{-3}}\right)^{1/3} \left(\dfrac{M_{\rm c}}{20 M_{\oplus}}\right)^{5/3} \left( \dfrac{r_{\rm p}}{1 \rm AU}\right) \left( \dfrac{5 M_{J}}{M_{p}}\right)^2
\label{eq:feedback}
\end{equation}

or alternatively,

\begin{equation}
    \dfrac{E_{\rm fb}}{E_{\rm tot}} = \left( \dfrac{M_{\rm c}}{M_{\rm p}} \right)^2 \dfrac{ r_{\rm p}}{r_{\rm c}}.
    \label{eq:fb_simple}
\end{equation}

Let us consider a protoplanet with $M_{\rm p} = 5M_J$, $
r_{\rm p}$ = 1 AU, and a rocky core with mass $M_{\rm c}=M_\oplus$ and $\rho_{\rm c}$ = 5 gcm$^{-3}$. Equation \ref{eq:feedback} gives us a value of $\sim$ 0.008, far too small to disrupt the protoplanet. However, if we allow the core to grow to 20 $M_\oplus$ \citep[Jupiter's core mass is now believed to be $\sim 10 M_{\oplus}$;][]{WahlEtal17,Nettelmann17,GuillotEtal18}, we get a value of 1.3. This result is exciting since it shows us that reasonably modest cores have the potential to destroy young GI protoplanets. 
Figure \ref{fig:an_feedback} illustrates Equation \ref{eq:feedback} for a 5 $M_J$ protoplanet with two different assumptions for the initial radius.
In the remainder of this paper we demonstrate the significance of this result using 3D simulations.

One of the simplifications of this argument is that we have neglected cooling inside the protoplanet. However, \cite{HelledBodenheimer11} found that the cooling timescale for protoplanets is typically in the range $10^4-10^5$ years, this tells us that if solid cores can grow faster than the cooling timescale for GI protoplanets then they have the potential to completely unbind them.

\section{Numerical methods}
\label{sec:methods}

This work extends previous simulations by \cite{NayakshinCha13}, \cite{Nayakshin17a} and \cite{HumphriesNayakshin18} and complements 3D simulations of core formation inside GI planets by \cite{Nayakshin18}.
We model gas and dust in protoplanetary discs with the coupled smoothed particle hydrodynamics (SPH) and N-body code \textsc{gadget-3} \citep[see][]{Springel05}.
Gas is modelled with an ideal equation of state with adiabatic index $\gamma = 7/5$, appropriate for a pre-collapse GI protoplanet dominated by molecular Hydrogen at a few hundred Kelvin \citep{BoleyEtal07}. An N-body tree algorithm is used to calculate the gravitational forces for all components in the system. We use a slight modification of the simple $\beta$ cooling model \citep{Gammie01} in which gas specific internal energy $u$ evolves according to

\begin{equation}
\dfrac{\mathrm{d}u}{\mathrm{d}t} = - \dfrac{u - u_{\rm eq}}{ t_{\rm cool}}\;,
\label{eq:beta0}
\end{equation}

where $u_{\rm eq} = k_B T_{\rm eq}/(\mu (\gamma-1))$, $k_B$ is the Boltzmann constant and $\mu = 2.45 m_p$ is the mean molecular weight for gas of Solar composition. The equilibrium temperature mimics irradiation from the star and is given by $T_{\rm eq}$ = 20K (100AU/R)$^{1/2}$. Similar to \cite{Nayakshin17a}, we set the cooling time $t_{\rm cool}$ to
\begin{equation}
t_{\rm cool}(R) = \beta \Omega_K^{-1} \; \left(1 + \dfrac{\rho}{\rho_{\rm crit}}\right)^{5} \;,
\label{eq:beta_def}
\end{equation}
where $\beta = 5$ and $\Omega_K = (GM_*/R^3)^{1/2}$ is the local Keplerian angular frequency at $R$. The term in the brackets in Equation \ref{eq:beta_def} quenches radiative cooling at gas densities higher than $\rho_{\rm crit} = 2 \times10^{-11}$ gcm$^{-3}$ in order to capture the long cooling timescales inside GI protoplanets \citep{HelledBodenheimer11}. 
Dust grains in our model are treated as a set of Lagrangian particles embedded in the gas that experience aerodynamical drag \citep{LAandBate15b}. We consider a variety of grain sizes ($a$) but consider a fixed grain density ($\rho_a$) of 3 gcm$^{-3}$ to simulate silicate grains. We use a semi-implicit integration scheme that captures both short and long stopping time regimes and include the back-reaction on the gas \citep{HumphriesNayakshin18}. Self-gravity of both dust and gas particles is computed via N-body methods.

\subsection{Modelling the solid core}
\label{sec:core_modelling}
Sink particles are often introduced in hydrodynamical simulations to deal with gravitational collapse of a part of the system that can no longer be resolved \citep[e.g.,][]{Bate95}. For the problem at hand, we introduce a dust-only sink in the protoplanet centre. We set the sink radius $r_{\rm sink}$ to 0.03 AU, comparable to the minimum SPH smoothing length in the centre of the protoplanet. Dust particles that are within $r_{\rm sink}$ and gravitationally bound to the sink are accreted. The sink particle is introduced at $t=0$ and its initial mass is set to $10^{-3}\mearth$, which is small enough to not influence the initial dynamics of either gas or dust. Note that collapse of the gas component onto the sink does not occur in our simulations even when the sink grows very massive. This is because the radiative cooling timescale for gas very near the solid core is much longer than the duration of these simulations.

We assign the core a constant material density $\rho_{\rm c}=5$~g~cm$^{-3}$. This represents a constant density mix of silicates and iron with a similar composition to the Earth ($\rho_{\oplus}$ = 5.51 gcm$^{-3}$). As this core grows, its gravitational potential energy is released into the gas at the centre of the protoplanet. If the energy release was instantaneous the accretion luminosity $L_{\rm c} = G \dot{M} M_{\rm c} /r_{\rm c}$, would be very high, since the rates of core growth within our protoplanets are as large as $\dot M_{\rm c} = 10^{-2} \mearth$~yr$^{-1}$.

However, the rate at which this potential energy is released is strongly dependent on the internal physics of the core and the core-gas boundary, neither of which we are able to resolve in this paper. \cite{BrouwersEtal18} show using 1D simulations in the context of the Core Accretion scenario that solids settling onto a massive core undergo vaporisation in the hot gas near to the core, this limits the rate at which feedback energy can be released during core growth. We set a parameter $t_{\rm acc}$ to smooth out energy release in order to account for this, although we are limited to relatively short time scales of $10^3$ and $10^4$ due to the numerical cost of our simulations. We address this point further in Section \ref{sec:discussion}.

In practice, the rate of the energy injection is therefore given by a luminosity $L_{\rm c}$ and follows the prescriptions from  \cite{NayakshinPower10,Nayakshin15b}. At any given time the luminosity of the core is given by

\begin{equation}
   L_{\rm c} = \dfrac{G M_{\rm c}}{r_{\rm c}} \left( \dfrac{M_{\rm res}}{t_{\rm acc}} \right)
   \label{eq:core_L}
\end{equation}

where $M_{\rm res}$ represents a `reservoir' of accreted pebbles that evolves such that 

\begin{equation}
    \dfrac{\mathrm{d} M_{\rm res}}{\mathrm{d} t} =  \dot{M}_{\rm peb} - \dfrac{M_{\rm res}}{t_{\rm acc}}.
\end{equation}

Note that this prescription releases the correct amount of energy when integrated over time and yields the instantaneous accretion luminosity in the limit $t_{\rm acc} \rightarrow 0$. 

Since we do not model radiative transfer in this paper, we adopt a simplified energy transfer prescription and inject the heat into the SPH neighbour particles of the sink.
The corresponding thermal energy is passed to the nearest 160 SPH neighbours of the sink, weighted by the SPH kernel\footnote{We vary this neighbour number in Appendix \ref{App:fb_NN}}.

\subsection{Simulation setup}
\label{sec:setup}

Our protoplanetary disc has a surface density profile of $\Sigma \propto R^{-1}$ (consistent with simulations of protoplanetary disc formation by \cite{Bate18}), an initial mass of $100 \mj$ and an outer radius of 100 AU\footnote{GI discs may be much larger, but we take 100 AU in order to reduce the numerical cost of the simulations}. The disc is relaxed for $\sim$ ten orbits at the outer edge. We then add dust particles to the disc at a suppressed vertical height of 10\% relative to the gas scale height in order to represent dust settling. After this, a $5 \mj$ metal-free protoplanet (modelled as a polytropic sphere with an initial radius of 3 AU) is injected into the gas disc on a circular orbit at 50 AU. The mass of the protoplanet ($M_{\rm P}$) is taken to be the mass within its half Hill-sphere.

For these simulations we assume that large grains dominate the mass budget of metals in the disc and therefore we set the initial disc metallicity to $Z_0=0.01$, such that the initial mass in pebbles is 1\% of the initial gaseous disc mass. This corresponds to 300$M_{\oplus}$ of metals, in line with the \cite{TychoniecEtal18} result for Class 0 discs. 
In \cite{HumphriesNayakshin18} we found that gas giants accreted nearly 100 percent of pebbles that entered their Hill spheres for a broad range of grain sizes from 0.03-30 cm. The rapid migration of the gas giants allowed accretion to proceed far beyond the `isolation mass' \citep[e.g.,][]{LambrechtsJ12}.

\subsection{Dust sedimentation inside protoplanets}
\label{sec:dust_sedimentation}
\begin{figure}
\includegraphics[width=1.0\columnwidth]{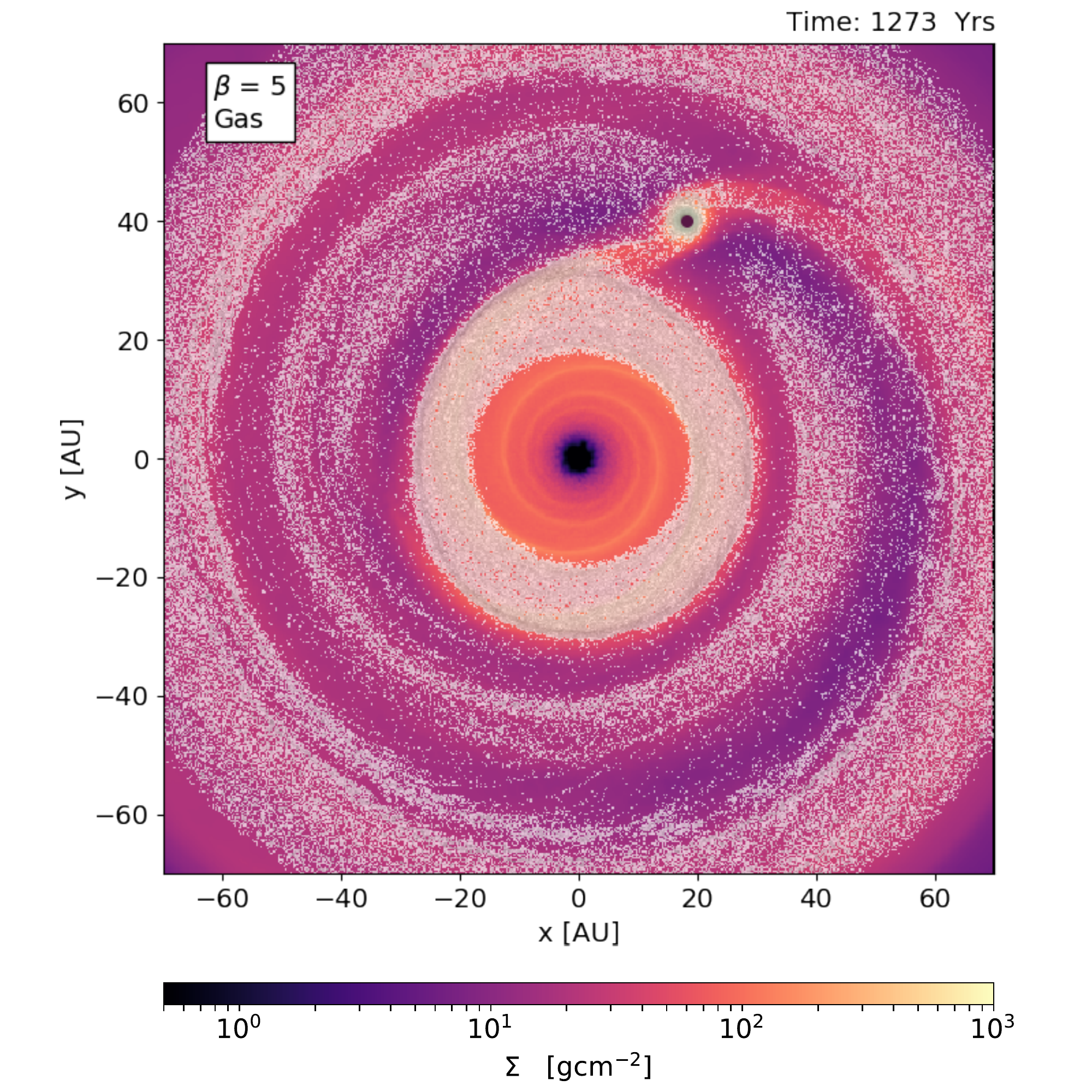}
\caption{Global gas surface density for a protoplanet with no feedback in a simulation with 1 cm pebbles. The pebbles are plotted with over-plotted greyscale points. The location of the protoplanet is marked with a black dot. Notice that the protoplanet has already carved a deep gap in the dust distribution in fewer than four orbits.}
\label{fig:no_fb_global}
\end{figure}

\begin{figure}
\includegraphics[width=0.99\columnwidth]{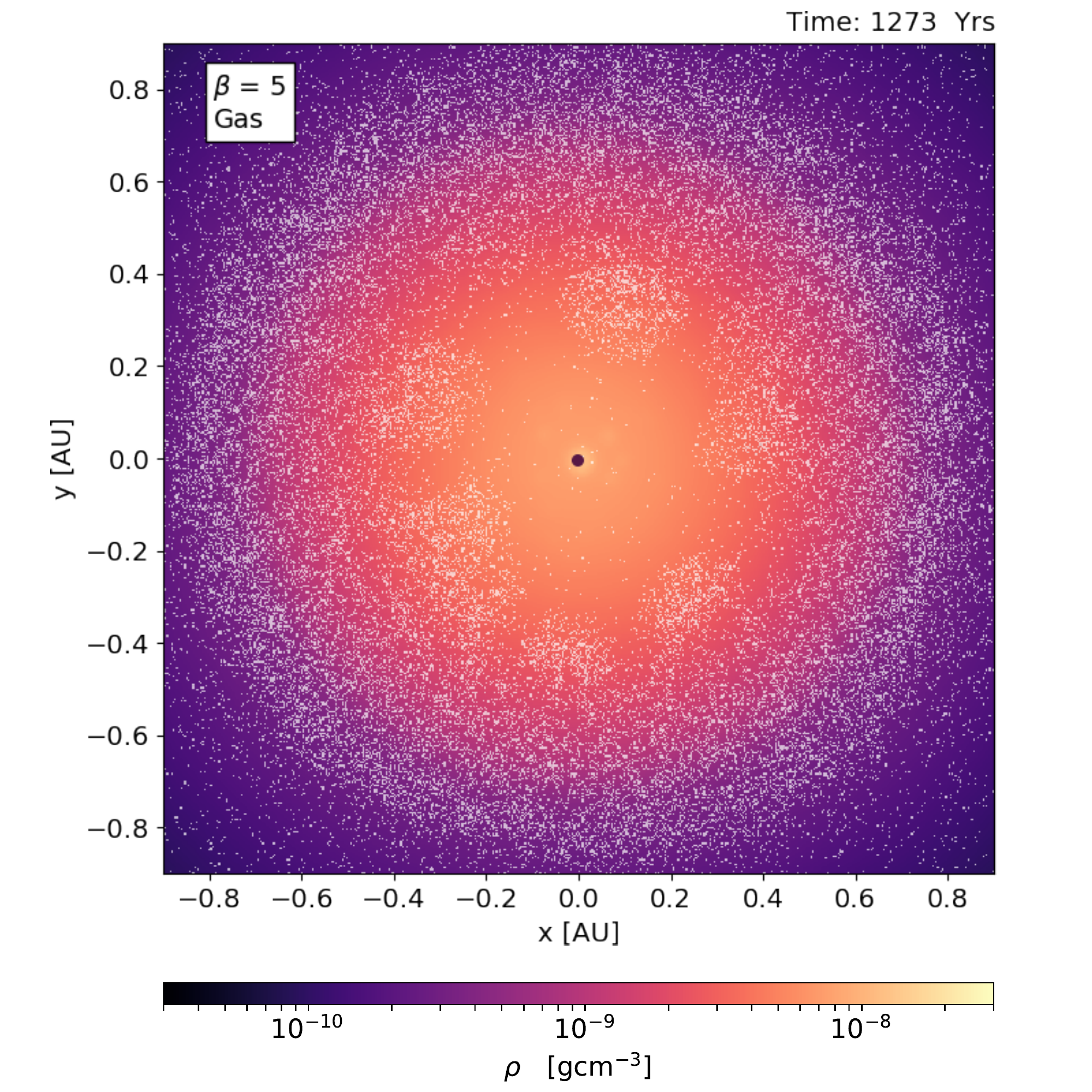}
\caption{A zoom view on the protoplanet from the same simulation as Figure \ref{fig:no_fb_global}, the core sink particle is marked with a black dot. The sedimentation time for 1 cm grains is thousands of years and so the majority of accreted grains remain suspended in the upper atmosphere of the protoplanet between 0.4-0.8 AU from the centre.}
\label{fig:no_fb_zoom}
\end{figure}

\begin{figure}
\includegraphics[width=1.0\columnwidth]{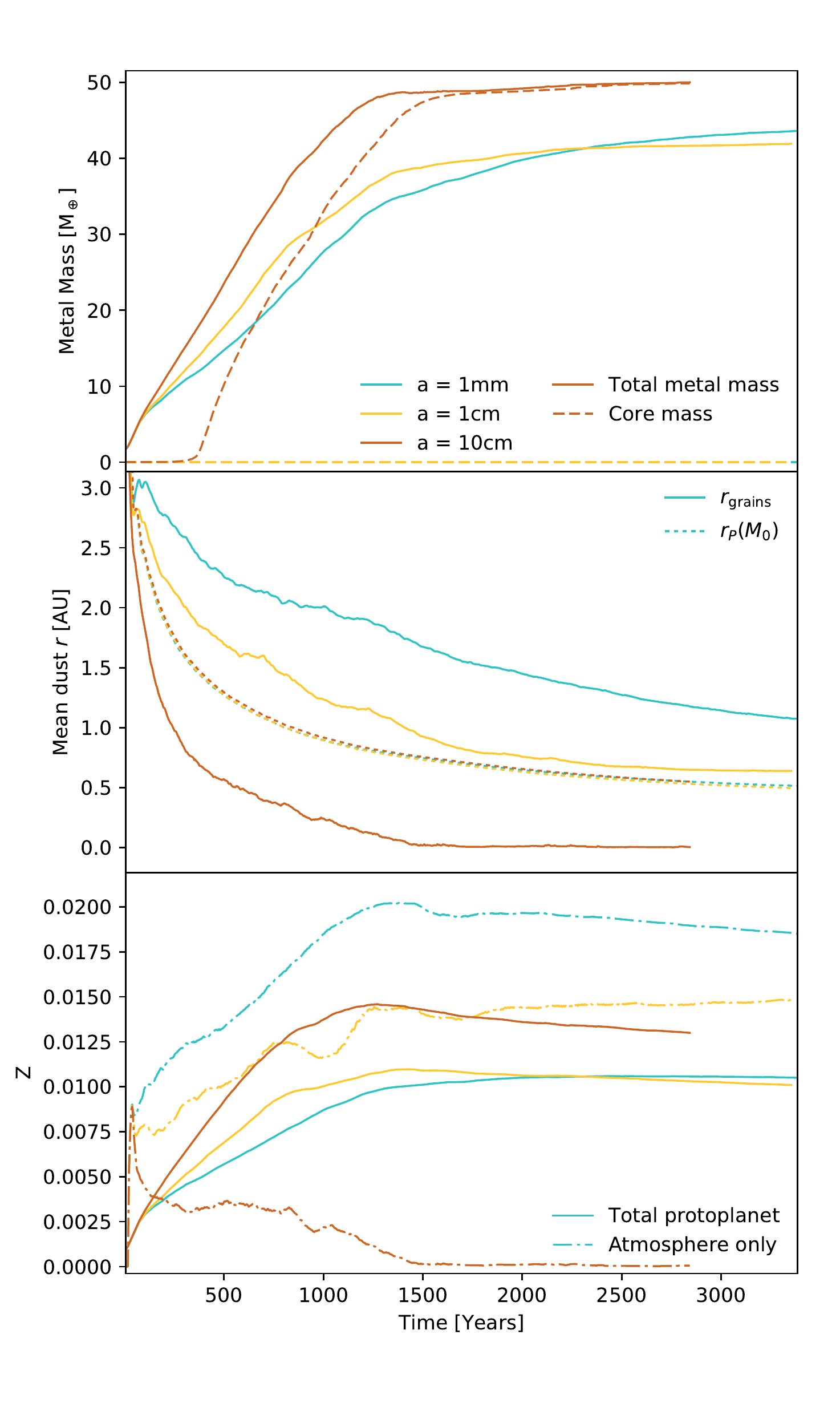}
\caption{Dynamics inside the protoplanet for different grain sizes. Top: Pebble accretion rates onto the protoplanet are comparable for mm, cm \& 10 cm grains \protect\citep{HumphriesNayakshin18}. Cores only grow for decoupled 10 cm grains while 1 mm and 1 cm grains remain in the atmosphere. Middle: Mean radial distance of dust from dust sink. Note the rapid settling of $a=10$cm grains. The dotted line shows the contraction of radius that contains the initial mass of the protoplanet ($r_P(M_0)$). Bottom: total protoplanet and atmosphere only metallic compositions. Small grains remain suspended in the atmosphere, causing it to become metal enriched.}
\label{fig:grain_stats}
\end{figure}

Before including feedback in our simulations, we explore the behaviour of differently sized dust species inside our protoplanets using the simulation setup described in Section \ref{sec:setup}. These simulations are identical to those in Section \ref{sec:fb}, save that core feedback is disabled. 

Figure \ref{fig:no_fb_global} shows the surface density of the gas disc for a simulation with 1 cm grains after 1273 years, the protoplanet core is marked with a black dot and is orbiting at 50 AU in an anti-clockwise direction. The location of pebbles in the disc is marked by the over-laid grey colour map. The protoplanet has carved a deep gap in the grains in less than four orbits. In agreement with our previous work \citep{HumphriesNayakshin18}, this shows that pebble accretion is very efficient for GI protoplanets. Figure \ref{fig:no_fb_zoom} shows a zoom view of the gas density inside the protoplanet in the plane of the disc from the simulation in Figure \ref{fig:no_fb_global}. Recall that pre-collapse protoplanets are initially very extended with radii of \jh{up to} a few AU. The core is marked with a black dot in the centre of the image while dust particles are marked in white. For the parameters of this protoplanet, 1 cm grains are expected to sediment on timescales of a few thousand years and so the majority remain suspended between 0.4-0.8 AU away from the centre of the protoplanet.

We now examine these simulations in more detail for a range of grain sizes, $a=0.1$~cm, $a=1$~cm and $a=10$~cm, which roughly correspond to Stokes numbers of 0.04, 0.4 and 4 in the disc.
The top panel of Figure \ref{fig:grain_stats} shows that the total mass of pebbles accreted from the disc inside the half Hill sphere of the protoplanet is very similar for all of the three grain sizes. This result is consistent with \cite{HumphriesNayakshin18} who found that pebble capture from the disc into the Hill sphere is almost 100\% efficient down to Stokes numbers of 0.1. Despite similarities in the accretion rates however, the dashed line shows that a core is only able to grow for the 10 cm grains.

The middle panel of Figure \ref{fig:grain_stats} shows the mean radial distance of the grains inside the protoplanet with respect to its centre (defined as the location of maximum gas density). 10 cm grains sediment to the core of the protoplanet rapidly whereas 1 mm and 1 cm grains remain suspended in the upper atmosphere of the protoplanet. This occurs because the sedimentation time of particles inside protoplanets depends linearly on grain size ($a$) as 

\begin{equation}
t_{\rm sed} = \dfrac{v_{\rm th}(r) \rho_{\rm g}(r) r^3}{M_{\rm enc}(r) G a \rho_a}
\label{eq:vsed}
\end{equation}

where $r$ is the distance from the protoplanet centre, $M_{\rm enc}$ is the enclosed mass, $\rho_a$ is the pebble density, $v_{\rm th}$ is the thermal gas speed and $\rho_{\rm g}$ is the gas density. For a typical protoplanet this gives a sedimentation timescale of $\sim 10^2$ years for 10 cm grains but $10^4$ years for mm grains. 

Since the protoplanet is embedded in the disc, it also accretes gas onto its atmosphere during the course of the simulation, causing the original protoplanet to contract. We now examine whether grains remain coupled to this new atmosphere or sediment inside the radius that contains the initial mass of the protoplanet ($r_P(M_0)$), which we plot as a dotted line in the middle panel of Figure \ref{fig:grain_stats}.
The orange line shows that only 10 cm grains are able to sediment inside this boundary, smaller grains remain suspended in the upper atmosphere of the protoplanet. Essentially, small grains remain coupled to the accreted disc material, whilst 10 cm grains penetrate into the protoplanet.

The bottom panel shows the fractional metallic composition of the total protoplanet and of only the atmosphere material. 1 mm grains cause a significant enhancement to the metallicity of the atmosphere, though in our simulations this enhancement is concentrated around the protoplanet midplane due to vertical dust settling in the global disc.

Based on Figure \ref{fig:grain_stats}, we choose to continue this study using only 10 cm grains. This choice is numerically convenient as simulating the slowly sedimenting 1 mm grains requires very long integration times that makes it impractical for 3D simulations. From the top panel we see that the total accretion rate onto the protoplanet is comparable for 1 mm and 10 cm grains, so our choice of grain size does not affect the mass budget of grains accreted onto the protoplanet. \cite{Nayakshin18} showed that grain growth time scales inside the protoplanet can be very short (as little as $10^2$ years), we therefore make the assumption that accreted grains rapidly grow to large sizes inside the protoplanet. Whilst a more thorough model of grain growth would provide a more reliable result, we are satisfied that this choice strikes a balance between numerical convenience and reasonable physical assumptions. One dimensional analysis of core growth for smaller grains can be found in \cite{Nayakshin16a}.

\section{Key result: protoplanet disruption}
\label{sec:results}

\begin{table}
\centering
\begin{tabular}{ c c c c c } \
\setlength{\tabcolsep}{0pt}
Name & $t_{\rm acc}$ [years]  & $Z_{\rm 0}$ & $N_{\rm disc}$ [$10^6$] & $N_{\rm PP}$ [$10^6$]\\
\hline
No Feedback               & -             & 0.01    & 2  &  0.1\\
$t_{\rm acc} = 10^3$ Yrs  & 1000      & 0.01    & 2  &  0.1\\
$t_{\rm acc} = 10^4$ Yrs  & 10,000    & 0.01    & 2  &  0.1\\
$Z_0$ = 0.3\%             & 1000      & 0.003   & 2  &  0.1\\
$Z_0$ = 3\%               & 1000      & 0.03    & 2  &  0.1\\
High res.                 & 1000      & 0.01    & 16 &  0.8\\

\hline
\end{tabular}
\caption{Summary of run parameters for the models presented in Figure \ref{fig:feedback}. $N_{\rm disc}$ and $N_{\rm PP}$ represent the total number of SPH particles in the initial disc and protoplanet respectively, there is typically one dust particle for every three SPH particles.}
\label{table:fb_params}
\end{table}

We now extend the simulations from the previous section in order to explore the effects of core feedback from the release of gravitational potential energy, using the prescriptions detailed in Section \ref{sec:core_modelling}. We find that in several cases this feedback is able to completely unbind young GI protoplanets. See Table \ref{table:fb_params} for a summary of the various models.

\subsection{Feedback from dust protocores}
\label{sec:fb}

\begin{figure}
\includegraphics[width=1.0\columnwidth]{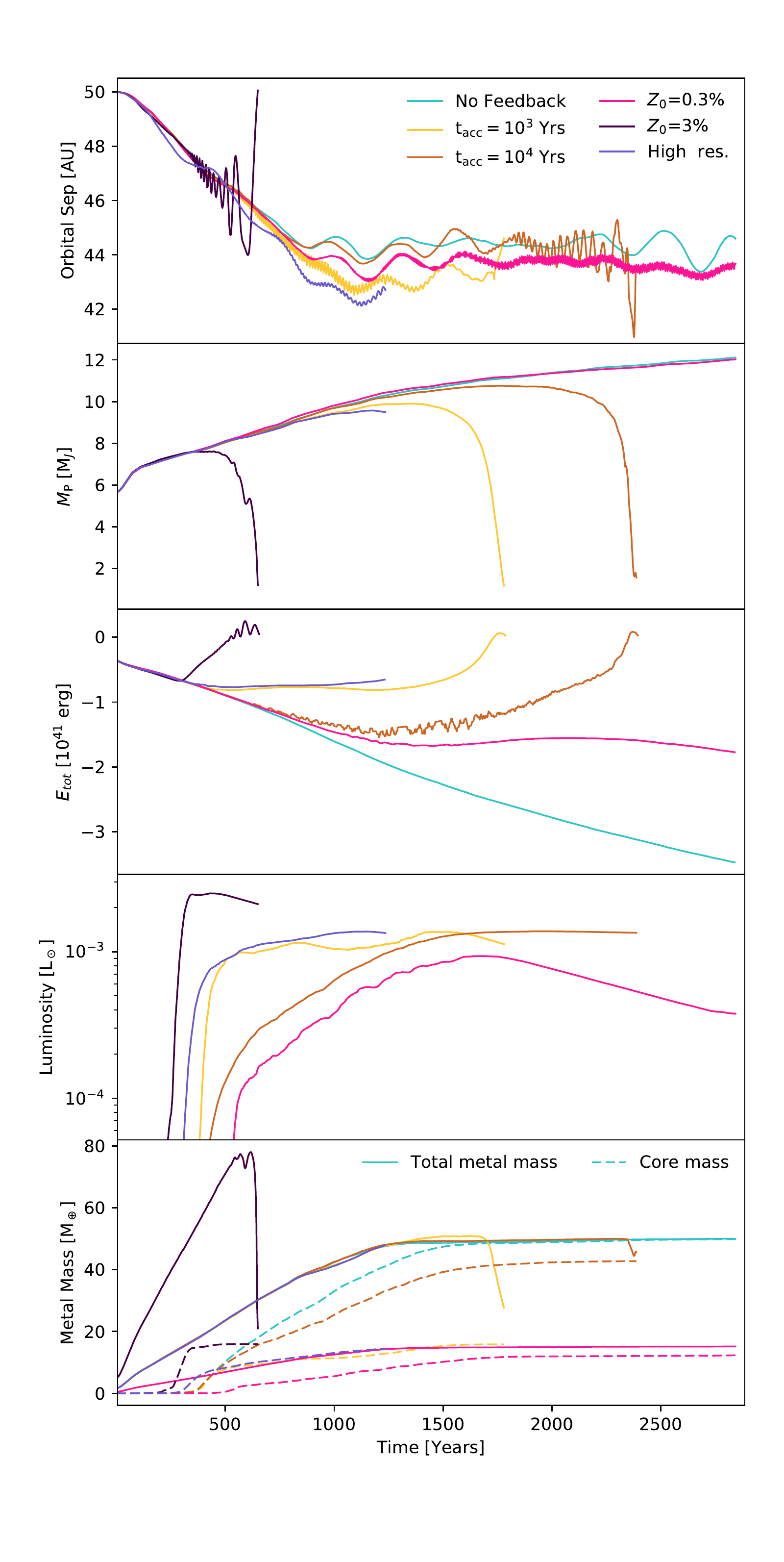}
\caption{The effect of including feedback from solid cores built through dust accretion. Top panel: Orbital separation of each core, cores are released on eccentric orbits after protoplanet disruption. Upper middle panel: Gas mass inside the half Hill sphere of each protoplanet. Middle panel: sum of thermal and gravitational potential energy of gas within the half Hill sphere. Lower middle panel: core luminosity; for this protoplanet the core must reach almost $10^{-3} L_{\odot}$ to cause a disruption.  Bottom panel: the solid lines plot the total metal mass inside the half Hill sphere. Dashed lines plot only the core mass.}
\label{fig:feedback}
\end{figure}

The top panel of Figure \ref{fig:feedback} charts the orbital distance between the star and the dust only sink particle for several simulations with a variety of initial dust to gas ratios and feedback timescales.
In all cases, the planet migrates inwards rapidly in the type I regime as expected from previous simulations \citep{BaruteauEtal11,Nayakshin17a} and begins to open a gap since it is relatively massive \citep{MalikEtal15,FletcherEtal19}.
The upper middle panel shows the mass inside the half Hill sphere of the protoplanet. The planet mass increases from 5 to 10 $\mj$ due to gas accretion, unless a disruption event triggered by core feedback takes place.  Notice that the final stages of disruption are always rapid, as expected from analytical work on the Roche lobe overflow of polytropic spheres with adiabatic index $\gamma = 7/5$ \citep{NayakshinLodato12}.

The middle panel shows the sum of the thermal and gravitational potential energy for gas in the protoplanet. In the no feedback case (cyan line) the protoplanet contracts due to external gas pressure from the disc and becomes increasingly more bound. When feedback is included, the internal energy of the protoplanet is modified considerably. 
The lower middle panel in Figure \ref{fig:feedback} plots the core luminosity as described in Section \ref{sec:core_modelling}.
If the core becomes sufficiently luminous then it will begin to increase the internal energy of its parent protoplanet, this leads to a disruption event once the total energy becomes positive. 

The solid lines in the bottom panel of Figure \ref{fig:feedback} show the total metal mass inside the protoplanet whilst the dashed lines show only the mass of the dust sink. With no feedback, the sink accretes all of the available 10 cm pebbles within 1500 years. The resulting core is 50$M_{\oplus}$, around 16\% of the initial metal mass in our simulation. 
The interplay between the accretion timescale and the available metal mass sets the final core mass at disruption. With a short feedback timescale of $10^3$ years we find a final core mass of 15 $M_{\oplus}$. However, if the feedback time is long the core is able grow to larger sizes before the feedback energy disrupts the protoplanet. 
For the fiducial parameters chosen in this paper ($5M_J$ protoplanet at $50$ AU, 10 cm grains and pebble to gas ratio of 1\%) we see disruption of our protoplanets when we set the accretion feedback timescale to $10^3$ and $10^4$ years. We also see a very rapid disruption when we set the initial pebble to gas ratio at $Z=$ 3\%. Setting $Z=$ 0.3\% suppresses the total available pebble mass, preventing a disruption event.
This shows crudely that disruption events are less likely in low metallicity environments.
After disruption, accreted grains not incorporated into the core are redistributed to the disc. This restocks the reservoir of large grains for pebble accretion onto subsequent protoplanets. In these simulations, disruption events release between ten and forty Earth masses of metals back into the disc. 

Our setup is conservative in that we assume a massive and initially metal free protoplanet in order to reduce uncertainty in the initial conditions; in reality we expect a factor of two initial metal enhancement of the protoplanet \citep{BoleyDurisen10}. 
The fact that disruption occurs in this conservative case demonstrates that core growth feedback provides a powerful mechanism for destroying GI protoplanets\footnote{Additional simulations not presented in this paper found that increasing the initial protoplanet metallicity makes disruption more efficient.}. 
On the other hand, our large grains and short feedback timescales (choices made for numerical convenience), help to speed up the disruption. We address these assumptions further in Section \ref{sec:discussion}.

\subsection{Inside the protoplanets}

\begin{figure}
\includegraphics[width=0.99\columnwidth]{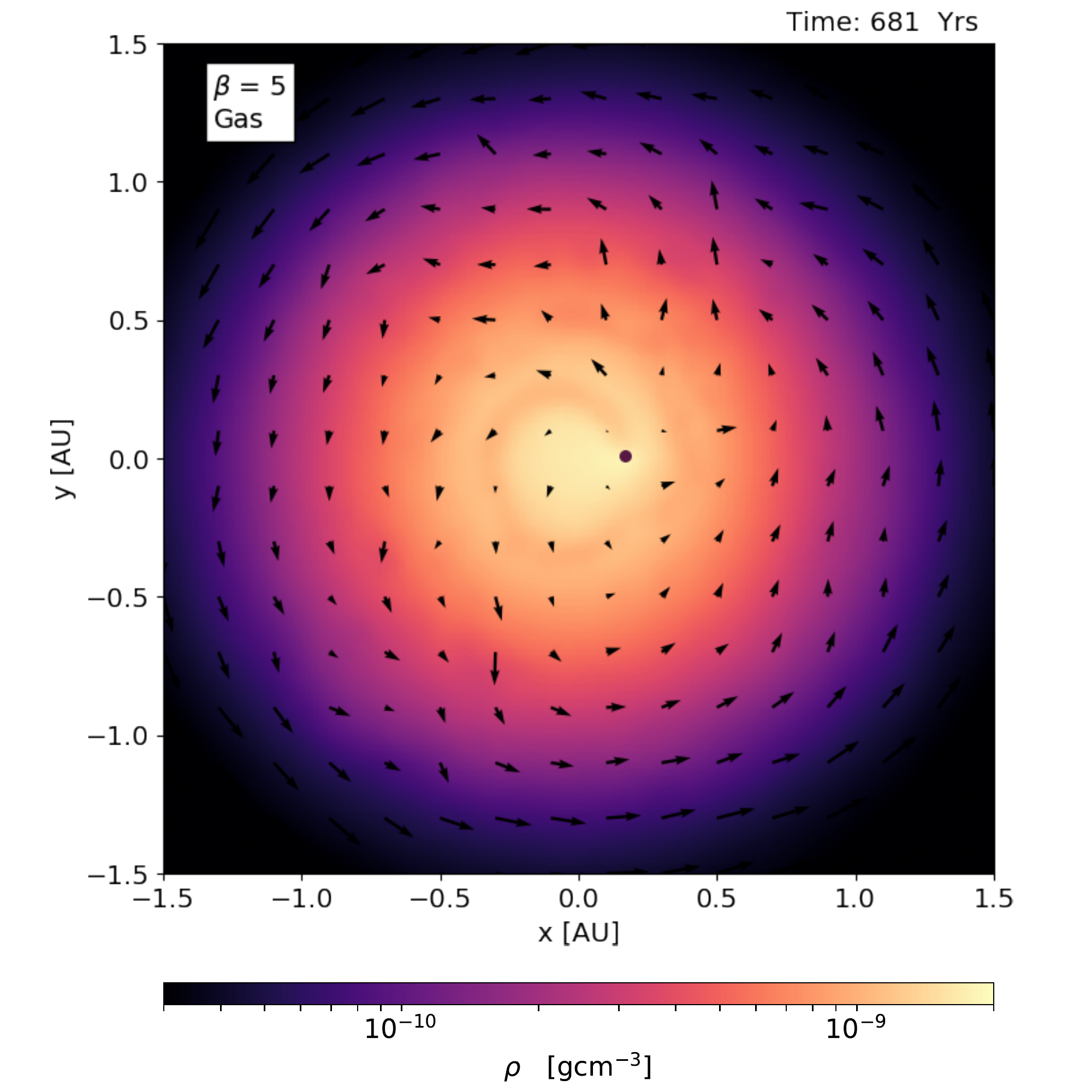}
\caption{Gas density inside the $t_{\rm{acc}}$=1000 years protoplanet in the x-y plane,  the sink particle is marked with a black dot. Note the weak rotational profile.}
\label{fig:zoom_rho_xy}
\end{figure}

\begin{figure}
\includegraphics[width=0.99\columnwidth]{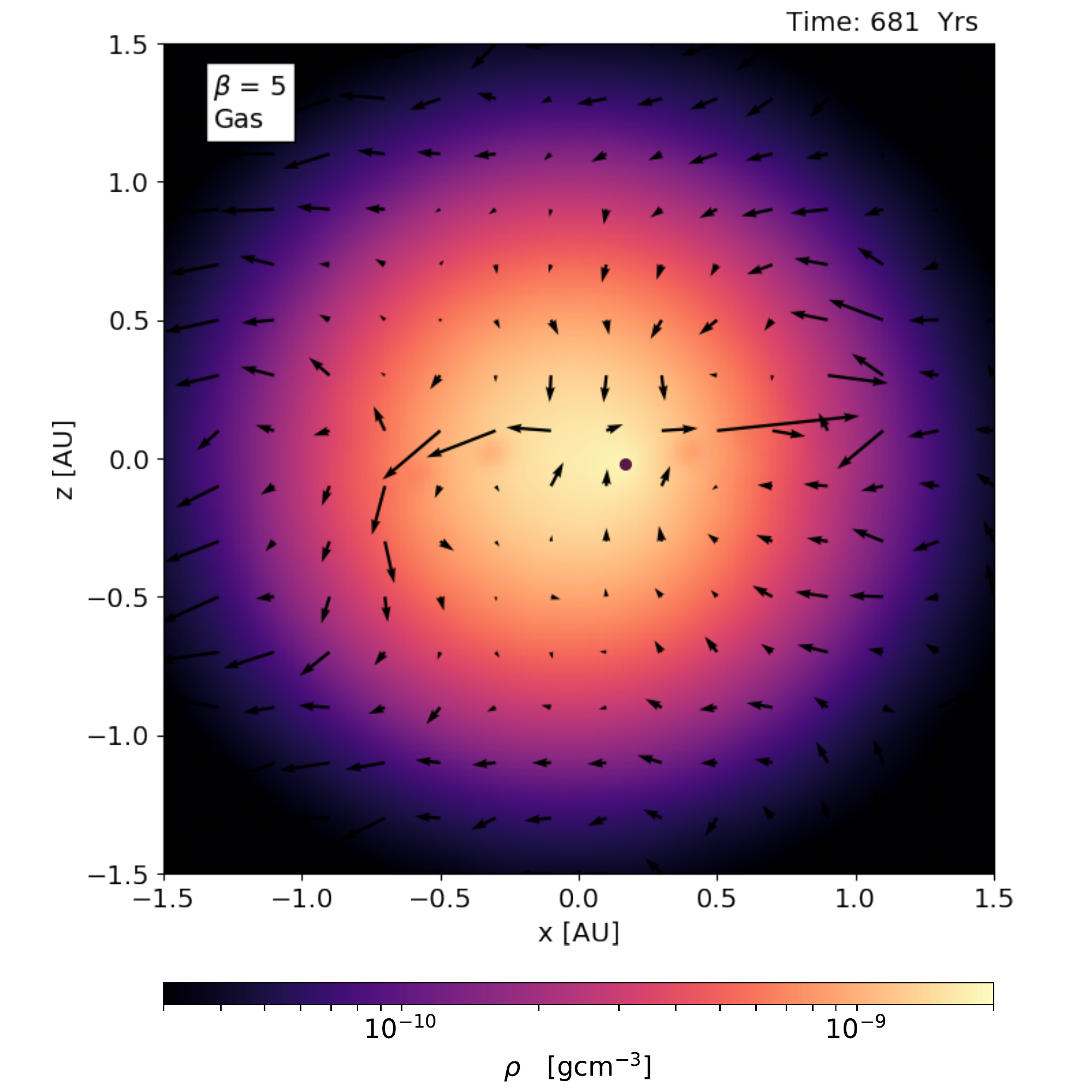}
\caption{Gas density inside the $t_{\rm{acc}}$=1000 years protoplanet in the x-z plane, the sink particle is marked with a black dot. The velocity field shows the outflow of hot gas through the midplane of the protoplanet.}
\label{fig:zoom_rho_xz}
\end{figure}

\begin{figure}
\includegraphics[width=0.99\columnwidth]{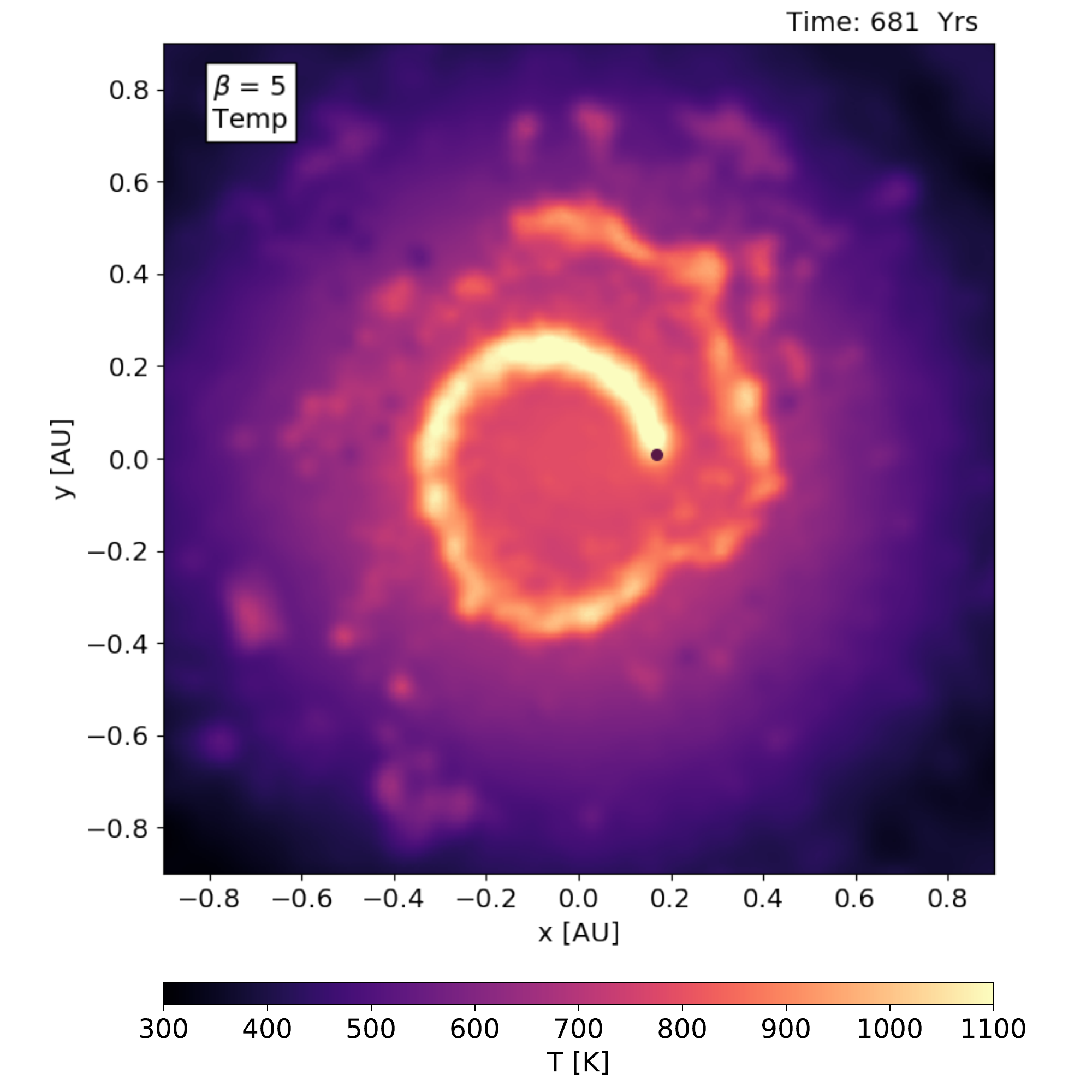}
\caption{Gas temperature inside the $t_{\rm{acc}}$=1000 years protoplanet. The sink particle is marked with a black dot and is orbiting in a clockwise direction with a period $\sim$20 years. Note the high temperature trail behind it. Escaping hot gas can be seen in an arc in the top half of the plot.}
\label{fig:zoom_T}
\end{figure}

Figures \ref{fig:zoom_rho_xy}-\ref{fig:zoom_T} show different views of the protoplanet from the $t_{acc}=1000$ years simulation in Figure \ref{fig:feedback} at 681 years. Figures \ref{fig:zoom_rho_xy} and \ref{fig:zoom_rho_xz} show density slices from this protoplanet in the x-y and x-z planes. We can see that  the density rises from $10^{-11}$ to $10^{-9}$ g~cm$^{-3}$  between the outer and inner regions. The overplotted black arrows in each figure show the velocity field. Figure \ref{fig:zoom_rho_xz} shows that low density gas from the core heated via feedback is escaping through the midplane. 
Figure \ref{fig:zoom_T} shows a zoom view of internal temperature, centred on the maximum central density of the protoplanet. Including core feedback heats gas in the centre of the protoplanet: once this heating becomes slightly anisotropic, hot gas begins to escape preferentially through low density channels and bubbles up to the surface. 
This process forms a low density gas trail behind the core as seen in Figure \ref{fig:zoom_T}. In this figure the core is orbiting in a clockwise direction with a period of $\sim$ 20 years.

It is immediately obvious from these figures that the core is moving inside the protoplanet, but what is causing this? In all of our feedback simulations we see that the core gradually leaves the minimum of the gravitational potential well of the protoplanet and begins to orbit this central region.

\section{Dynamics of the luminous core}
\label{sec:core}

\begin{figure}
\includegraphics[width=0.99\columnwidth]{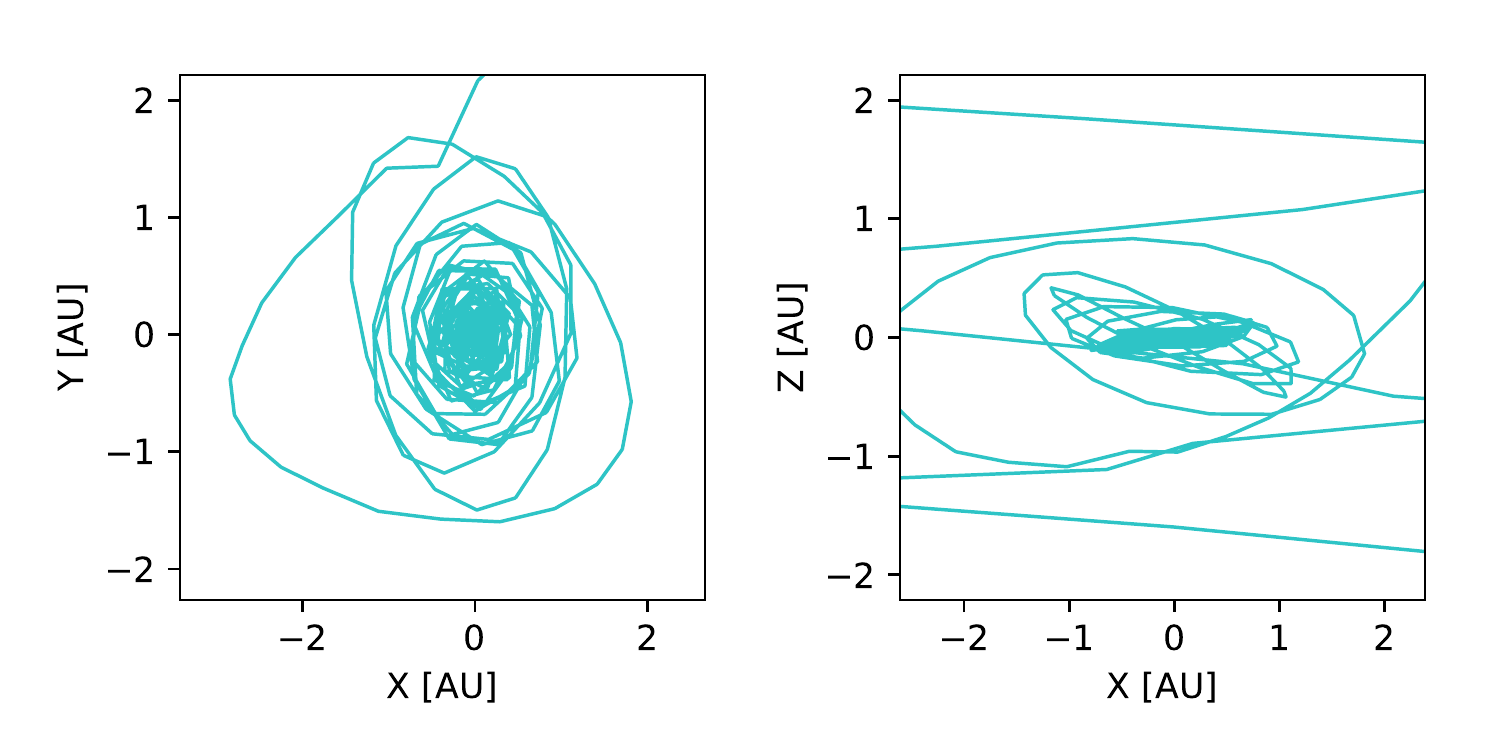}
\caption{Tracks of the core motion for the $t_{acc}$ = 1000 years simulation in the x-y and x-z planes, centred on the maximum gas density inside the protoplanet. The core starts at the centre then spirals outwards due to the heating force, but mostly remains in the x-y plane. After disruption the core is released with a slight eccentricity}
\label{fig:core_tracks}
\end{figure}

In Figure \ref{fig:core_tracks} we plot tracks to show the motion of the core in the x-y and x-z planes for the $t_{acc}$ = 1000 years simulation. We see that once the core leaves the central $\sim 0.2$~AU region its orbit appears to be rather circular. We also note that the orbit of the core is not confined to the x-y plane; since our feedback scheme is homogeneous and the centre of the protoplanet is essentially spherically symmetric it is not obvious that there should be a preferential orientation. In this simulation the core reaches an offset radius of $\sim$ 1 AU with respect to the centre of the protoplanet before the disruption event happens at $\sim$ 1700 years.

Similar, seemingly bizarre, `wandering' behaviour for massive luminous objects has recently been found by authors in different branches of Astrophysics.
The displacement of luminous super-massive black holes from the centre of a young gas-rich galaxy was found by \cite{SijackiEtal10} in their simulations of merging super-massive black holes displaced from the galactic centre due to gravitational kicks. They found that luminous black holes were ejected further than expected given their kick and did not come back to the galaxy centre afterwards. Black holes of zero luminosity, on the other hand, returned to the galaxy centre in agreement with classical dynamical friction theory. \cite{SijackiEtal10} identified the asymmetry in the gas distribution around the SMBH as the driver of this unexpected behavior. In Chandrasekhar dynamical friction, there is a higher density trail {\em behind} the massive perturber. In the case of a very luminous object, a high temperature, low density halo inflated by the perturber is blown off by the higher density headwind, and thus the trail behind the perturber is a low density one. Therefore, the direction of the gravitational torque on the perturber changes sign and acts to accelerate it. \cite{ParkBogdanovic17} found a similar effect in simulations of a SMBH producing ionizing feedback on the background neutral medium.

\cite{MassetVelascoRomero17} performed a detailed analytic study of this {\em heating force} exerted on a massive object due to a hot tail of gas in an otherwise homogeneous gaseous medium.  In particular, they equate the dynamical friction force ($\vec{F}_{df}$) with the heating force ($\vec{F}_{heat}$) to find an equilibrium speed $V_0$ at which the object moves through the medium. In the sub-sonic limit,
\begin{equation}
V_0 = \dfrac{3 \gamma (\gamma-1) L_{\rm c} c_s}{8 \pi \rho_0 G M_{\rm c} \chi},
\label{eq:v_MVR17}
\end{equation}
where $\gamma$ is the adiabatic index of the gas, $L_{\rm c}$ and $M_{\rm c}$ are the luminosity and mass of the perturbing core and $\chi$ is the thermal conductivity of the gas that they assumed to be constant. They noted that the effects of the heating force are expected to be significant for planet formation application, via, e.g., the Earth developing a non-negligible eccentricity and inclination with respect to the protoplanetary disc for realistic disc parameters. These conclusions and the analytic result (Equation \ref{eq:v_MVR17}) were confirmed with numerical simulations by \cite{VelascoRomeroMasset19}, \cite{ChrenkoLambrechts19} and \cite{GuileraEtal19}.

In application to our particular problem, we note that the density in the central part of the protoplanet is initially homogeneous with a nearly constant $\rho_0 \approx 10^{-9}$~gcm$^{-3}$. Hence, the constant background density results of \cite{MassetVelascoRomero17} are applicable. However, our protoplanet is finite in extent. When the core is displaced by a distance $R$ from the centre, there is a returning force of protoplanet gravity, given by
\begin{equation}
    F_{\rm g} = - \frac{G M(R) M_{\rm c}}{R^2}\;,
    \label{eq:Fg0}
\end{equation}{}
where $M(R) \approx (4\pi/3) \rho_0 R^3$ is the enclosed mass within radius $R$.

The motion and radius of the orbiting core is then set by a balance of forces. In the azimuthal direction the heating and dynamical friction force specify the core velocity from Equation \ref{eq:v_MVR17}. Given this velocity ($V_0$), the radial position of the core ($R$) is simply set by the gravitational force of the enclosed protoplanet which provides the centrifugal force.
We see that $V_0$ is independent of orbit radius $R$ whereas the gravitational force is propotional to $R$, as long as the enclosed gas density is roughly constant. The balance of these therefore establishes an equilibrium radius of the orbit for the core of mass $M_{\rm c}$ and luminosity $L_{\rm c}$: 

\begin{equation}
    R_0 = \left(\frac{3}{4\pi G\rho_0}\right)^{3/2} \frac{\gamma (\gamma-1) c_s}{2\chi} \frac{L_{\rm c}}{M_{\rm c}}\;.
    \label{eq:R0}
\end{equation}{}

\begin{figure}
\includegraphics[width=1.0\columnwidth]{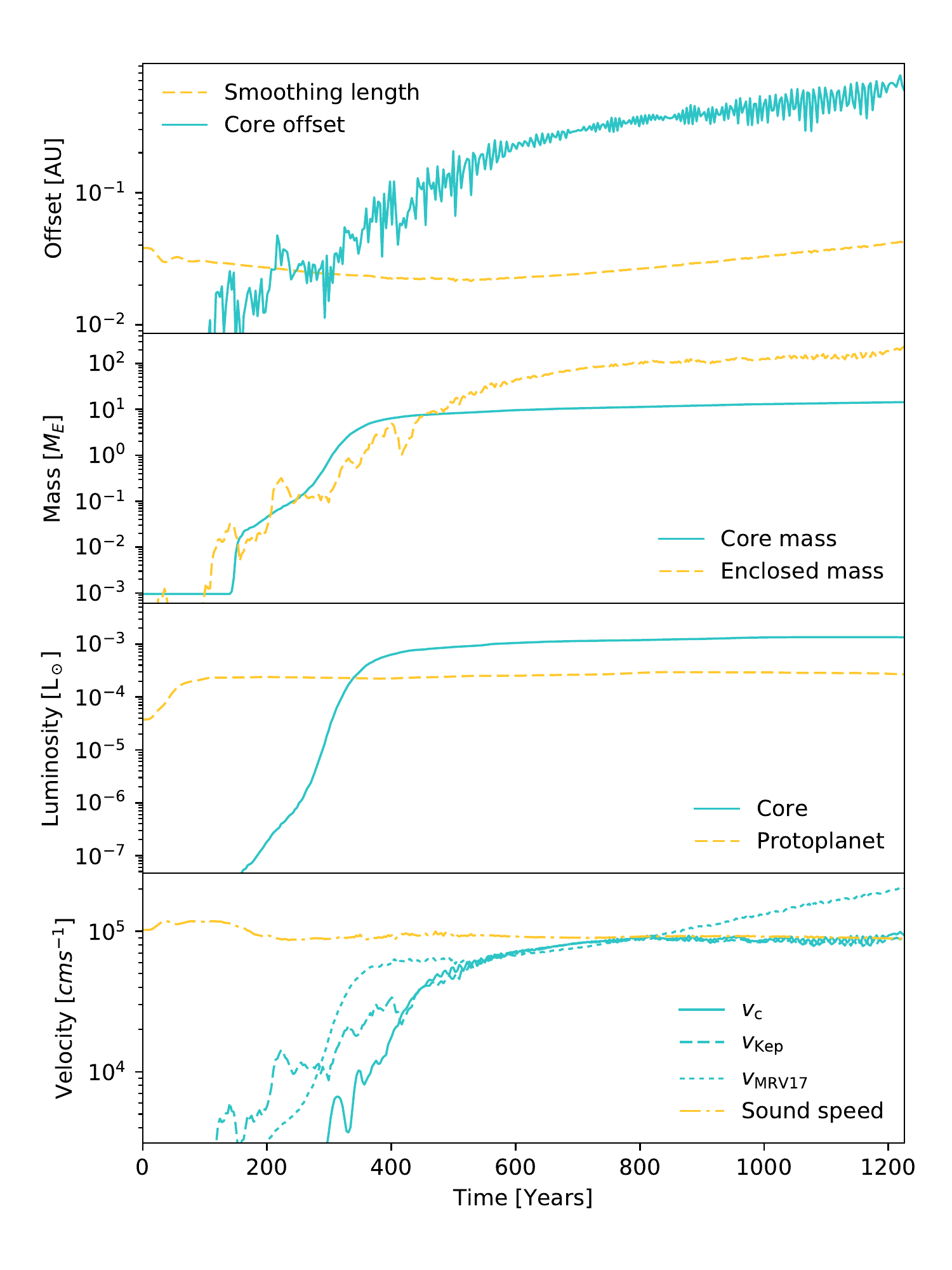}
\caption{Core diagnostics for the high res. $t_{\rm{acc}}$=1000 years protoplanet. Top: core offset from protoplanet centre (solid) and central SPH smoothing length (dashed). Top middle: core mass and mass enclosed between the offset core and the protoplanet centre. Bottom middle: core luminosity due to pebble accretion driven growth. Bottom: velocity of the core (solid), Keplerian velocity based on enclosed mass (dashed), predicted core velocity based on thermal acceleration \citep[dotted,][]{MassetVelascoRomero17} and central protoplanet sound speed (yellow).}
\label{fig:fb_vel}
\end{figure}

The \cite{MassetVelascoRomero17} problem setup assumes that there is a heat flow through the gas via thermal diffusion with a diffusivity coefficient $\chi$. However, our current models do not include thermal diffusion. In our simulations the heat is instead transferred from the core to the surrounding gas using the prescription described in Section \ref{sec:methods} above. The diffusivity of the heat flow from the core into the surrounding gas can then be estimated through a simple dimensional analysis as
\begin{equation}
    \chi = h c_{\rm s}\;, 
    \label{chi0}
\end{equation}{}
where $h$ is the SPH smoothing length inside the protoplanet. We emphasize that this estimate reflects the heat flow in our numerical feedback implementation. In Appendix \ref{App:fb_NN} we demonstrate the dependence of our results on numerical resolution, which affect $h$, and we also discuss a more physical estimate for $\chi$. 

With this assumption about $\chi$, Figure \ref{fig:fb_vel} compares the motion of the core with the \cite{VelascoRomeroMasset19} theory. The top panel shows the offset radius of the core relative to the maximum gas density inside the protoplanet. The dashed line plots the smoothing length inside the protoplanet which marks the SPH resolution limit inside the protoplanet. We see that over time feedback causes the core to drift further and further from the central density, the core has clearly left the central 0.1 AU of the protoplanet only a few hundred years after the onset of feedback.
The second panel shows the mass of the core (solid) and the mass of enclosed gas (dashed) within the offset radius. We see that the enclosed mass dominates the core mass once the core offset grows above 0.1 AU. The third panel shows the luminosity of the core calculated from Equation \ref{eq:E_fb} (also seen in Figure \ref{fig:feedback}), we see that this is $\sim 10^{-3} L_{\odot}$ once the core reaches $\sim 10 M_{\oplus}$. The yellow dashed line also plots the luminosity of the protoplanet, this is computed by summing the rate of energy loss for each SPH particle inside the protoplanet using Equation \ref{eq:beta0}. Despite our simplified cooling scheme, this luminosity agrees very well with the luminosity of isolated protoplanet calculations from \cite{VazanHelled12} who calculated that a 7 $M_J$ protoplanet has a luminosity of $\sim 2 \times 10^{-4} L_{\odot}$ for its first ten thousand years of life.

The bottom panel of Figure \ref{fig:fb_vel} is the most useful for understanding this process. We plot the velocity of the core (solid), the predicted Keplerian velocity for the core based on an orbit around the enclosed mass (dashed), the predicted \cite{MassetVelascoRomero17} velocity from Equation \ref{eq:v_MVR17} (dotted) and the sound speed in the gas (yellow). 
We see that the velocity of the core is very close to the Keplerian velocity, demonstrating that once it has moved beyond 0.1 AU it is orbiting the central gas mass of the protoplanet. We also see that the velocity of the core seems to be limited by the sound speed inside the protoplanet.
We do not expect the analytic theory to apply if (a) the core offset $R_0$ is smaller than $h$; (b) The enclosed mass within $R_0$ is smaller than $M_{\rm c}$. Finally, the agreement of the theory and simulations breaks down as $V_0$ approaches the sound speed. This could be because the core orbit starts to close-in on itself as it orbits around: the hot tail isn't completely dispersed by the time the planet makes one full orbit.
The assumption of constant background density also breaks down away from the centre of the protoplanet.

\subsection{A second core and smaller debris formation}\label{sec:2nd_core}
\begin{figure}
\includegraphics[width=0.99\columnwidth]{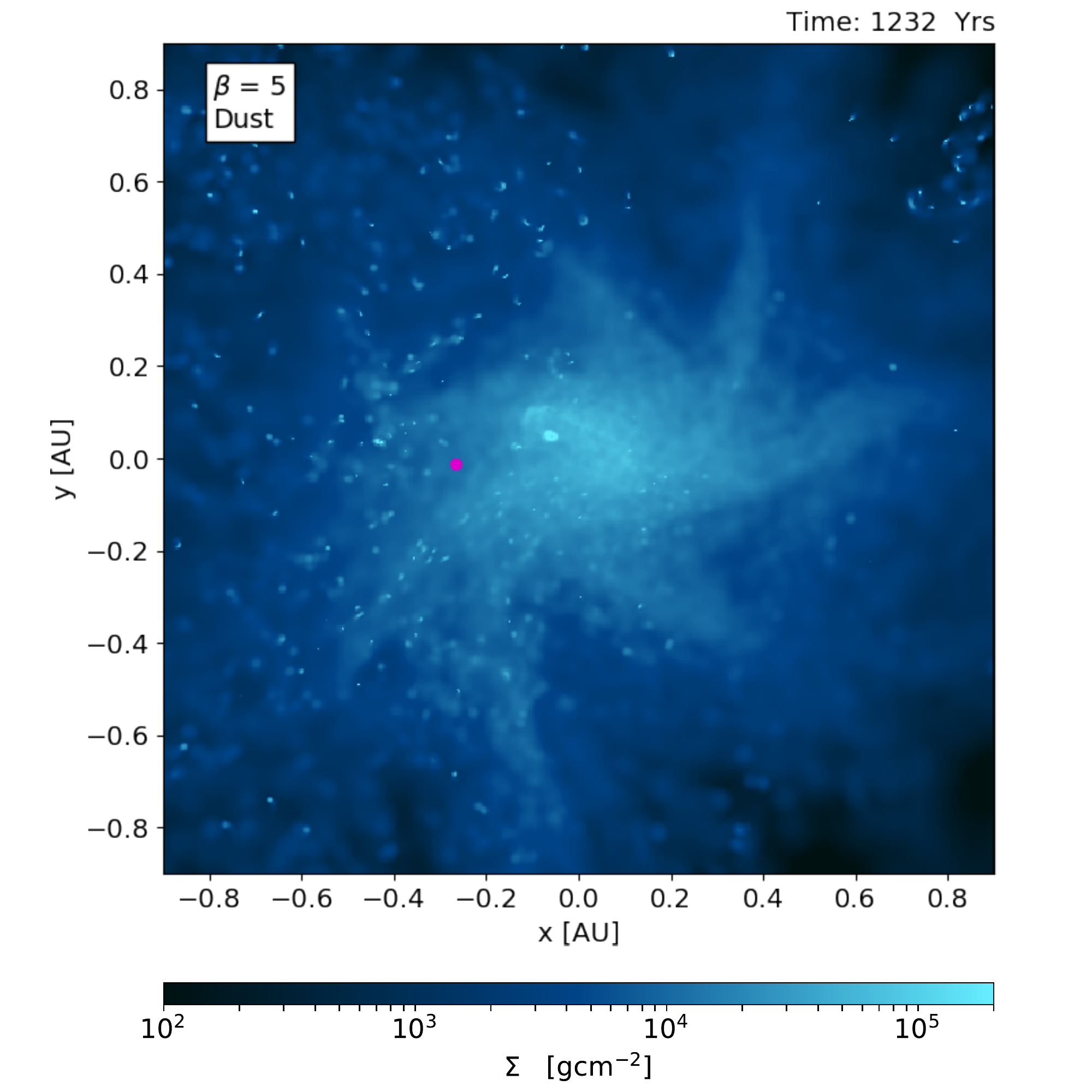}
\caption{Dust density slice from the 2x res run, centred on the maximum gas density. The core is marked by the pink dot and is orbiting at 0.4 AU. Dust continues to collect at the protoplanet centre, eventually causing the simulation to stall. There is approximately an Earth mass of solids in the central 0.04 AU, potentially sufficient to start a second core. This feature can also be seen in the lower resolution runs.}
\label{fig:zoom_dust}
\end{figure}

\begin{figure}
\includegraphics[width=0.99\columnwidth]{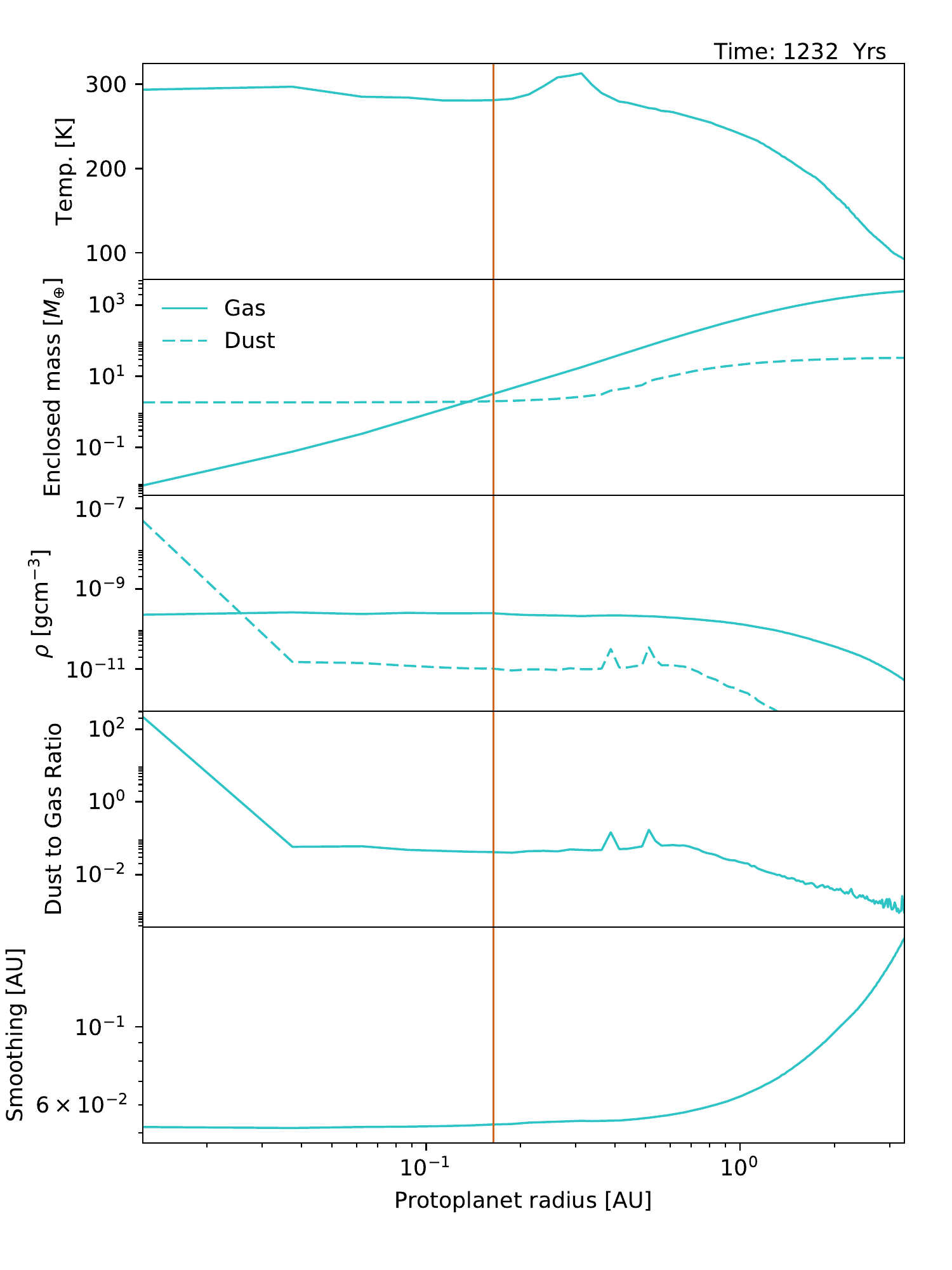}
\caption{Interior profiles for the high resolution protoplanet at 1232 years. The orange line indicates the radial position of the perturbed core at this time. The top panel shows internal temperature, since the gas is adiabatic the heating from the orbiting core does not reach the central regions. Panel two shows the enclosed dust and gas masses, 1.8 $M_{\oplus}$ of dust particles have collected at the centre of the protoplanet.
Panel three shows the enhancement of the central dust density due to sedimentation. Panel four shows that the innermost dust to gas ratio reaches a value of over 200, this has implications for secondary core formation. The bottom panel shows the SPH smoothing length. Note that the \textsc{gadget-3} smoothing length is defined to be twice as large as that commonly used in other SPH codes.}
\label{fig:zoom_profile}
\end{figure}

Once the core is driven away from the centre of the protoplanet, the rate at which it accretes pebbles drops significantly. Meanwhile, sedimenting pebbles continue to be focused to the centre of the protoplanet and collect there. Figure \ref{fig:zoom_dust} shows a plot of the dust surface density in the centre of the protoplanet from Figure \ref{fig:fb_vel}, the pink dot shows the location of the orbiting core. The core is orbiting at 0.3 AU from the centre of the protoplanet which has allowed a large mass of dust to collect in the central 0.1 AU.

Figure \ref{fig:zoom_profile} displays information about the central regions of our high resolution protoplanet at the same time as Figure \ref{fig:zoom_dust}, centred on the maximum dust density inside the protoplanet. 
The top panel shows the shell averaged gas temperature. Since cooling time in the centre of the clump is very long, the central gas is adiabatic, and thus thermal energy escapes in bubbles (as seen in Figure \ref{fig:zoom_T}). This leaves a slightly cooler zone of gas inside the orbit of the core. In addition, heating due to the core energy release is now also offset meaning that the peak temperature in the clump is no longer at its center. Of course, gas in the vicinity of the core may be much hotter than the shell averaged value plotted in Figure \ref{fig:zoom_profile}. In particular, Figure \ref{fig:zoom_T} shows that gas within 0.05 AU of the core reaches temperatures of 1200 Kelvin.

Panels two, three and four of Figure \ref{fig:zoom_profile} plot the enclosed gas and dust masses, the shell averaged gas and dust densities and the dust-to-gas ratio. They show that dust collects on scales below the SPH smoothing length in the centre of the protoplanet. 1.8 $M_{\oplus}$ of dust has collected at the centre of the protoplanet below the smoothing length resolution, this corresponds to a dust-to-gas ratio of over 200. In broader terms, the dust-to-gas ratio increases above one inside the inner 0.1 AU of the protoplanet.
Panel five shows the shell averaged SPH smoothing length\footnote{Note that the \textsc{gadget-3} smoothing length is defined to be twice as large as that commonly used in other SPH codes \citep{Springel05}.}. It reaches a value just over 0.05 AU in the central regions, which is sadly almost three orders of magnitudes greater than the radius of Earth. It remains very challenging to resolve core formation directly in 3D simulations of GI planet formation.

In Figure \ref{fig:zoom_dust} we also see numerous additional dust clumps of much smaller mass formed in this simulation. Two of the largest of these are also seen as small spikes in the dust density profile in Figure \ref{fig:zoom_profile} at 0.4 and 0.5 AU away from the central dust concentration. It is tempting to interpret this as evidence for formation of planetesimal-like bodies within gas clumps in the framework of the GI theory, as proposed by \cite{NayakshinCha12}, but we caution that the presence of these clumps and their properties depend on our numerical resolution parameters. Future high resolution 3D modelling will be necessary to correctly capture core and potential planetesimal formation inside GI protoplanets.

\section{Discussion}
\label{sec:discussion}

\subsection{Overview of core driven protoplanet disruption}

Our main results can be summarised as follows. Pebble accretion transports grains from the disc into the outer regions of a protoplanet. If the grains grow large enough (in our simulations $a=$10 cm) they sediment rapidly into the protoplanet and start to form a solid core. Now the outcomes diverge. If pebble accretion rates are high, the core is able to grow quickly and may be able to disrupt the protoplanet by over-heating and unbinding its gaseous envelope. 
Alternatively, when pebble supply is subdued, or grain growth is too slow, core growth is not vigorous enough to disrupt the protoplanet from within. What happens to the planet then does not depend on the core properties but depends on the competing effects of planetary migration versus planet contraction due to radiative cooling. The planet could be disrupted with only a small core surviving \citep{BoleyEtal10} or it may collapse and survive as a proper GI gas giant with a small core \citep{HelledEtal08}.  Similar conclusions were already reached in \cite{Nayakshin16a}; however in this study we have modelled the disruption of GI protoplanets by the core feedback for the first time in 3D SPH simulations.

Protoplanet disruption provides a promising mechanism for forming $\sim 10 \mearth$ cores and perhaps Saturn-mass planets (if gas envelope removal is only partial) at wide-orbits within the first $10^5$ years of disc evolution. Such objects are now invoked for explaining dozens of gaps and rings observed in dust emission of young discs \citep{Alma2015, AndrewsEtal16, LongEtal18, DSHARP1,DSHARP7}. We note that for our scenario to work, we require a reservoir of pebbles with a total mass of $\sim$ a hundred $M_{\oplus}$. This condition is satisfied by many of the observed ringed disc systems, although ALMA surveys are most sensitive to emission from $\sim 1$ mm sized dust, which is smaller than the $a=10$~cm used here. However, the discs with annular structures are older than our $t\approx 0$ disc, i.e., typically in the range of $1-10$~Myr. It is possible that these discs contained even more dust when they were younger, and that their first million years of life is sufficiently long to allow mm-sized grains to grow and sediment to form cores inside protoplanets on longer timescales than studied in this work \citep{HS08,Nayakshin10a,Nayakshin10b}. 

In contrast, forming wide-orbit super-Earth mass cores at such young ages is very challenging for the planetesimal-based Core Accretion theory \citep[e.g.,][]{KL99}. While pebble accretion may work much faster \citep{OrmelKlahr10,LambrechtsEtal14}, more recent work emphasised the inefficiency of locking pebbles into planets \citep{OrmelLiu18, LinEtal18}. 

Our results are particularly relevant given recent developments in the field. Meteorite population arguments suggest that Jupiter had already reached a mass of $20 M_{\oplus}$ within the first million years of the lifetime of our solar system \citep{KruijerEtal17}. Additionally, disc masses observed at one million years seem too low to explain observed planet masses, suggesting more evidence for rapid core formation to `hide' additional metal mass within the first million years of disc lifetime \citep{ManaraEtal18}.

\subsection{Connection to Hall et al. (2017)}

\begin{figure}
\includegraphics[width=0.99\columnwidth]{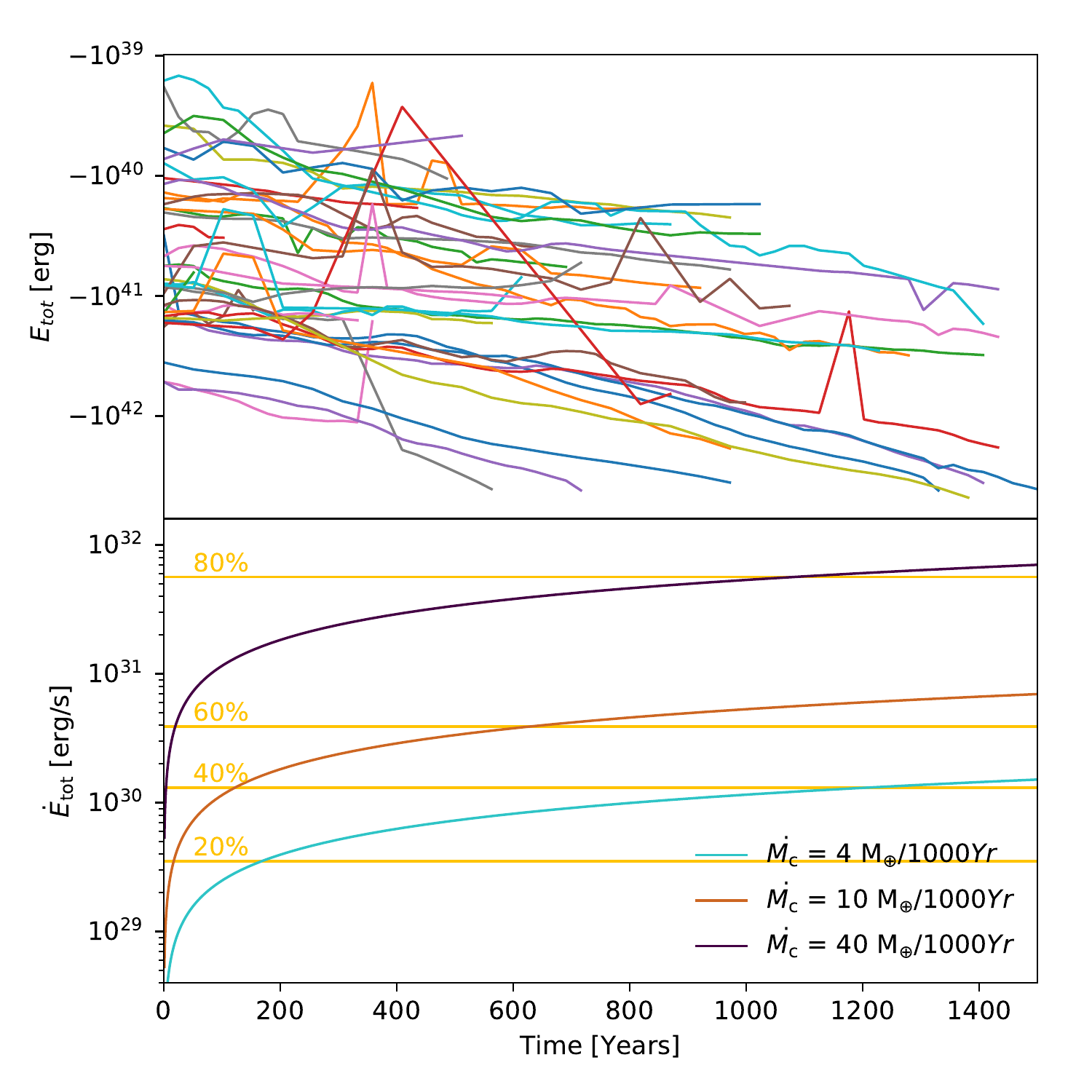}
\includegraphics[width=0.99\columnwidth]{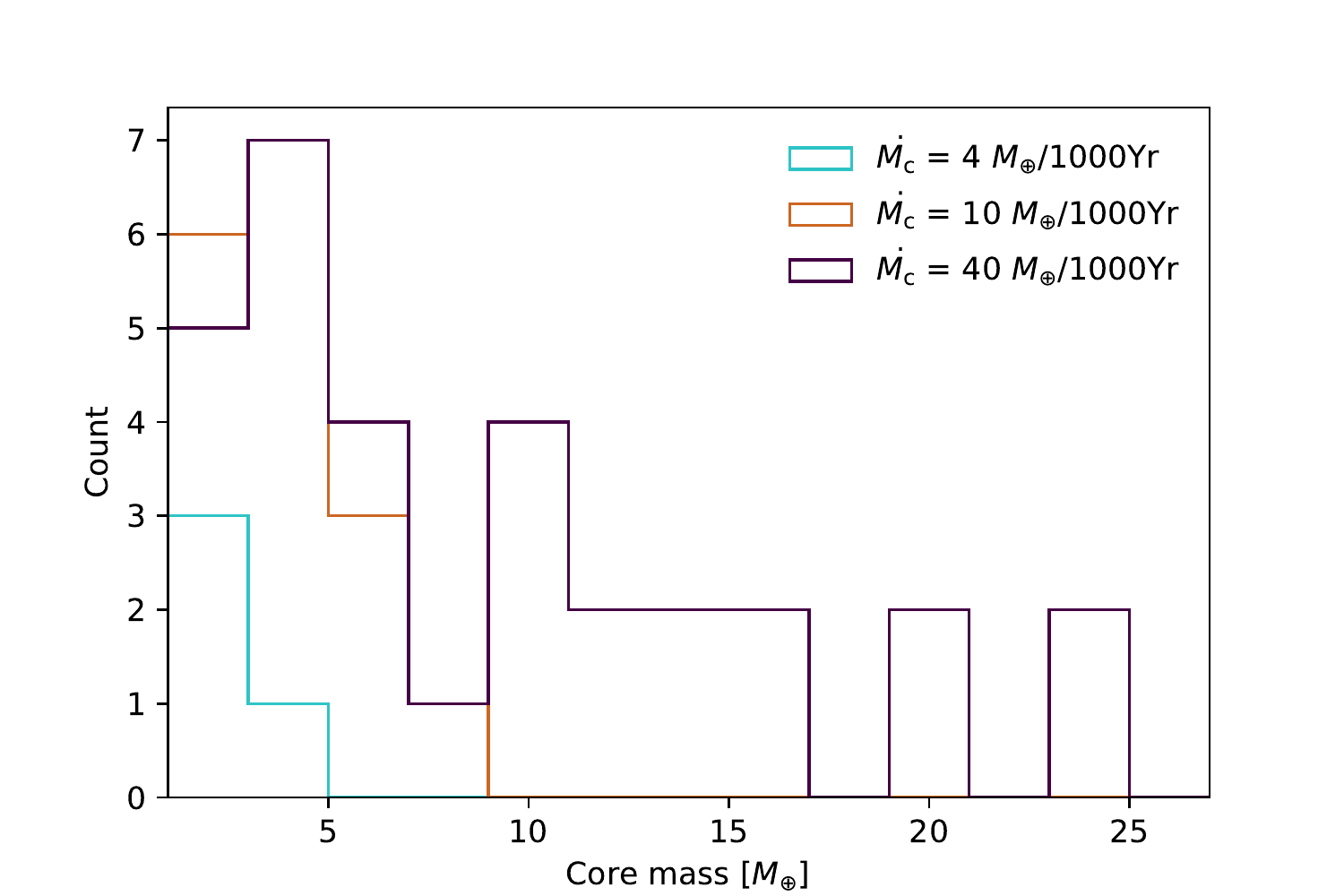}
\caption{Comparison to data from \protect\cite{HallEtal17}. Top: internal energy of the \protect \cite{HallEtal17} protoplanets against time. Middle: purple, orange and cyan lines mark core luminosity against time for different rates of pebble accretion driven core growth. The intersection of these lines with the yellow lines marks the point at which core luminosity dominates over the rate of internal energy decrease for 20, 40, 60 and 80 percent of the \protect\cite{HallEtal17} protoplanets. Bottom: histogram of final core masses in cases where the energy release during core formation was greater then the total energy of the protoplanet after only 1500 years.}
\label{fig:Cass}
\end{figure}

It is important to examine how our conclusions relate to the wider field of gravitational instability protoplanet formation. In order to do this we examine the internal energies of GI protoplanets from \cite[][hereafter H17]{HallEtal17} and compare them to our models. In contrast to our idealised radiative cooling prescription, the simulations of H17 use a more sophisticated approach to allow self-gravitating SPH discs to self-consistently fragment and form protoplanets. These protoplanets then evolve and contract until the simulation timestep in their cores becomes prohibitively small. 

Starting from the output of their simulations, we calculate the internal energies of these protoplanets and plot them in the top panel of Figure \ref{fig:Cass}. These show a broad spread in values but are generally comparable to  the typical protoplanet binding energy of $\sim$ 10$^{41}$ erg found in our simulations. In order to estimate whether core disruption may be a significant process for the H17 protoplanets, we calculate the average rate of change of their internal energies over time. We then take the analytic luminosity due to core formation used in Section \ref{sec:analytic_fb} and compare these two quantities. When they are equal, the collapse of the protoplanet will be first stalled and then reversed.

The middle panel of Figure \ref{fig:Cass} plots the core luminosity against time for three rates of core growth comparable to those we found in our 3D simulations (see Figure \ref{fig:feedback}). The intersection of the diagonal and horizontal yellow lines marks the time at which the core luminosity balances the rate of decrease in internal energy for 20, 40, 60 and 80 percent of the H17 protoplanets. Even under the smallest core growth rate considered, e.g., 4 $M_{\oplus}$ per 1000 years, core luminosity becomes dominant for 40 percent of the H17 protoplanets after 1000 years\footnote{These rates may seem high for core accretion, but they are motivated by high surface densities in young GI discs and the large capture radii of migrating Jupiter mass protoplanets.}. This percentage rises to 80 if core growth is very rapid.
These results indicate that if pebble accretion driven core growth is vigourous, it may be able to destroy many of these otherwise bound protoplanets within their first few thousand years of life. 

In the bottom panel of Figure \ref{fig:Cass} we plot a histogram of core masses that are able to unbind H17 protoplanets in less than 1500 years. Using Equation \ref{eq:feedback} we compared the total energy of the protoplanet with the total energy released due to core formation for the three different rates of core growth. The core masses are distributed between 1-25 $M_{\oplus}$ as expected from the analysis in Section \ref{sec:analytic_fb}, the sub $10 M_{\oplus}$ cores are formed since many of the H17 protoplanets initially have lower binding energies than our 5 $M_J$ protoplanet. It is likely that in a self-consistent calculation more of the H17 protoplanets would be at risk of disruption since core feedback slows the rate of protoplanet collapse and therefore allows more time for the growth of larger cores. Further analysis of this process will require self-consistent simulations of protoplanet evolution coupled with core feedback models, we direct the reader to \cite{Nayakshin16a} and (Nayakshin submitted) for 1D examples of this.

These results have major implications for the final predicted population of GI protoplanets, provided massive rocky cores form before the protoplanets undergo H$_2$ dissociative collapse. The release of energy has the potential to destroy many GI protoplanets, leaving behind massive solid cores in their place. Further studies are needed to quantify the subsequent fate of these planets.

\subsection{Super-Earth cores and other by-products of protoplanet disruption}

For the range of parameters considered in this paper, it seems that the critical core mass needed to trigger protoplanet disruption is roughly 5-30 $M_{\oplus}$. This is supported by the analytic estimates of core feedback in Section \ref{sec:analytic_fb}.
There are two mechanisms that limit core growth. First, if the feedback luminosity is high, protoplanet disruption happens early and pebble accretion onto the core ends.  In this case the subsequent core growth will be limited by the pebble isolation mass of $\sim 10-20\mearth$, studied extensively in the CA field\footnote{Note that \cite{HumphriesNayakshin18} show that this pebble isolation mass does not apply to the pre-disruption massive GI planets since they are rapidly migrating through the disc, and they sweep pebbles as they go.} \citep{MorbidelliNesvorny12,LambrechtsEtal14,BitschEtal18}.
Second, the total available mass of pebbles in the disc can limit core growth. This can be seen in Figure \ref{fig:feedback} for the $Z_0=0.3$\% and the no feedback runs. If there is no disruption, the protoplanet should carve a deep and wide gap in the global pebble distribution which should be observable within the first $10^5$ years.
We have sampled only the tip of the parameter space of this process; relative disruption rates will vary depending on the initial size and spatial distribution of pebbles; the mass and orbital separation of protoplanets and their interior density and temperature profiles.

In our simulations protoplanets are disrupted very quickly, e.g., at $t\ll 10^4$~yrs. It must be stressed that this result is strongly dependent on the size of pebbles and the internal structure of the protoplanet. In our simulations we used very large pebbles, $a=10$~cm, motivating this by the fact that grain growth inside the clump can be very rapid \citep[e.g.,][]{HelledEtal08,Nayakshin16a}. However, detailed models \citep[e.g., Figures 5-8 in][]{HB10} of planet evolution show that grain growth and the sedimentation process couple with convective cooling of the planet in very non-linear ways.

Furthermore, \cite{BrouwersEtal18} found for CA that once cores grow above 0.5 $M_{\oplus}$, pebbles sublimate before reaching the core surface and instead form a metal rich atmosphere. Based on this work, the rapid formation of a high density core may be an extreme assumption. However, our rates of pebble accretion are orders of magnitude higher than those considered in \cite{BrouwersEtal18} and it is unclear how this might affect their conclusions. Pebble sublimation would reduce the magnitude of gravitational potential energy released in our simulations, though calculation of the exact reduction will require an update to the \cite{BrouwersEtal18} models for GI cores. 
Additionally, the equation of state in the central regions of the protoplanet needs to be modified to include dust latent heat and other chemical processes ignored in our paper \citep{PodolakEtal88,BrouwersEtal18}. It is possible that the end product of a disruption event could be a partially disrupted object that still holds on to some of its primordial gas envelope. Better models resolving the region immediately surrounding the core are needed in order to properly constrain the end products of the core disruption process.
In light of this, the final masses of our cores remain model dependent.

As stated in the opening of this paper, we have assumed core formation in order to explore an extreme scenario for this problem. A less concentrated core would reduce the total available feedback energy, but would not necessarily prevent core feedback from disrupting less massive, initially metal enriched or more extended GI protoplanets. We have chosen to examine the extreme case of a rapid disruption with large pebbles and a short accretion time, in part due to the prohibitive computional cost of running our simulations for more than a few $10^3$ years. However, since most ALMA observations of protoplanetary discs are made at $\sim 10^6$ years the disruption process could be slower and still explain the presence of `young' cores. 
A potential cause of such a delay could be caused by less efficient cooling and contraction due to the high dust opacity in metal enriched protoplanets, coupled with a less extreme rate of core growth.
In fact, \cite{Nayakshin16a} studied similar scenarios for smaller grain sizes using 1D radiative transfer simulations and did indeed find that disruption events took place at later times. Those 1D simulations will be explored further in Nayakshin (submitted).

We also found that feedback causes these proto-cores to wander away from the protoplanet centre, allowing many Earth masses of dust to collect there in their absence. The formation of these `secondary cores' suggests that the by-products of core driven disruption events may be a rich collection of debris, in addition to the primary rocky core. We leave the study of this debris to future work due to the resolution limitations of our simulations.

Finally, the orbits of the super-Earth cores released back into the disc in our simulations are initially mildly eccentric and may be inclined by a few degrees to the disc midplane. This is caused by the chaotic nature of the feedback inside their spherically symmetric parent protoplanets; the orbit of the core within the protoplanet need not be confined to the plane of the global disc.
It is questionable whether these initial deviations from a circular orbit would survive the dissipation of the disc since \cite{TanakaWard04} showed that the eccentricity and inclination damping timescales are much shorter than the migration timescales for low mass planets.

\section{Conclusions}
\label{sec:conclusions}

Our 3D simulations support suggestions made on the basis of an earlier 1D study \citep{Nayakshin16a} that GI protoplanets can be disrupted by core driven feedback, thus reducing their mass at tens of AU. The mechanism is general enough to work at yet wider separations, and may be promising for forming, broadly speaking, Saturn mass and super-Earth ALMA planetary candidates.

Since feedback causes our rocky cores to drift from the protoplanet centre, we see tantalising hints that these disruption events could also lead to a rich array of secondary cores and debris. This may have important implications for the formation of giant planets with their own satellite systems.  Furthermore, the disruption of protoplanets with a rich retinue of smaller solid bodies may be an alternative way of forming debris discs and possibly the Kuiper belt in the Solar System \citep{NayakshinCha12}, but confirmation of this will require further and higher resolution modelling of dust and gas dynamics inside young protoplanets.

Although we did not focus on this topic specifically, we note that core-initiated disruptions of GI protoplanets may also help explain why they are so rarely observed at wide separations.
Many argue that GI rarely occurs since the fraction of wide-orbit companions above 2$M_J$ is now thought to be only $\sim$ 1\% \citep{ViganEtal17}. However, GI protoplanet formation can be much more prevalent if a large fraction of GI protoplanets are disrupted from within. Coupled with work demonstrating rapid protoplanet migration \citep{BaruteauEtal11,MullerEtal18}, it seems that the outcome of GI is not limited to massive wide-orbit planets. In fact, these objects way well be in the minority.

\section*{Acknowledgements}
We would like to thank the anonymous referee for their helpful comments and to thank Cass Hall for the output of simulations from \cite{HallEtal17} which we used in Section \ref{sec:discussion}. JH and SN acknowledge support from STFC grants ST/N504117/1 and ST/N000757/1, as well as the STFC DiRAC HPC Facility (grant ST/H00856X/1 and ST/K000373/1). DiRAC is part of the National E-Infrastructure.

\bibliographystyle{mnras}
\bibliography{humphries}


\appendix
\section{Varying the feedback injection lengthscale}
\label{App:fb_NN}

\begin{figure}
\includegraphics[width=1.0\columnwidth]{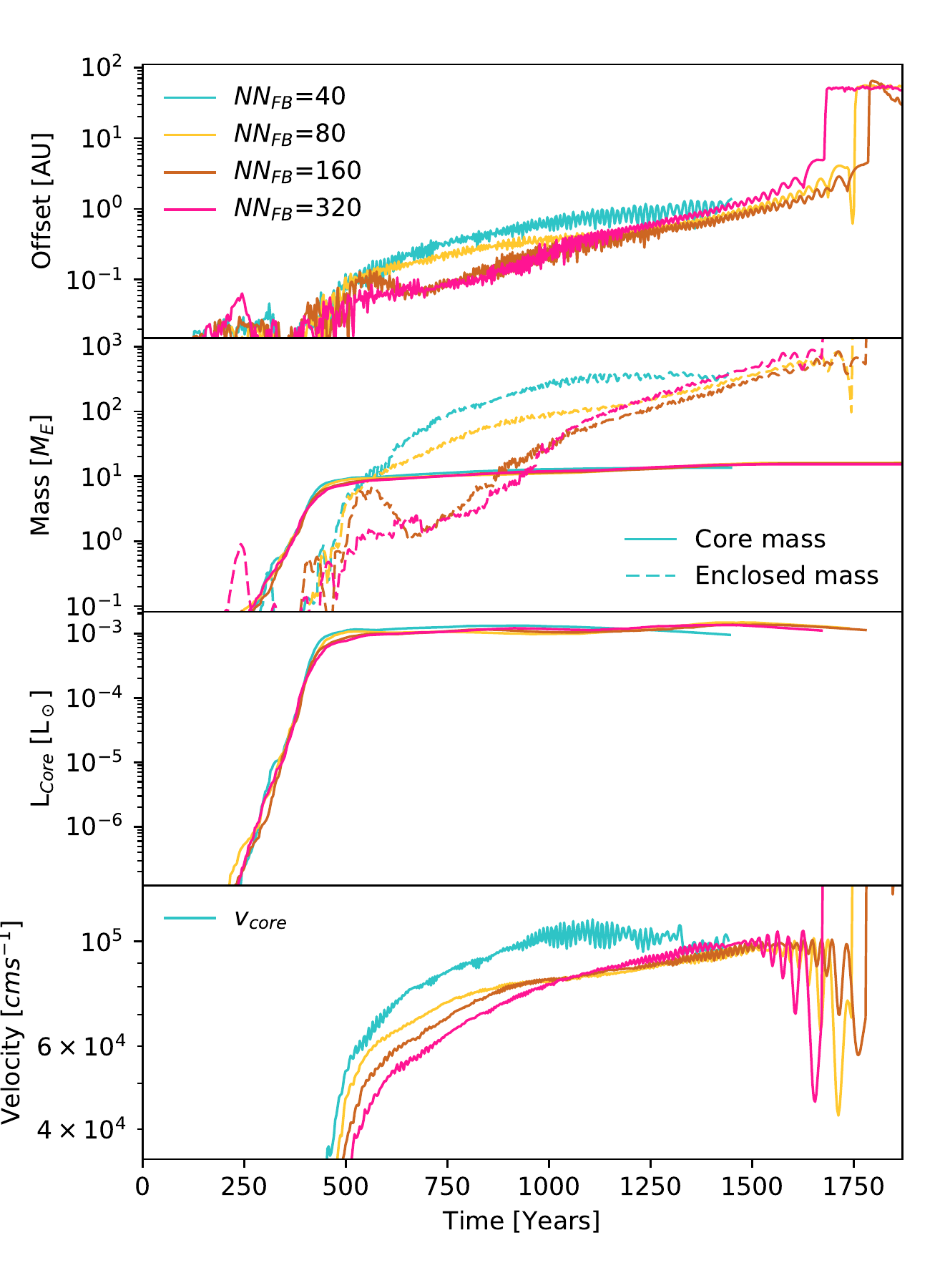}
\caption{Same as Figure \ref{fig:fb_vel}, but now for different feedback injection length scales. Whilst the feedback scheme changes the amount by which the core is offset from the centre of the protoplanet, the final disruption time remains comparable in each case. Final disruption is set by the total energy injected by the core regardless of its offset.}
\label{fig:fb_NN}
\end{figure}

\begin{figure}
\includegraphics[width=0.99\columnwidth]{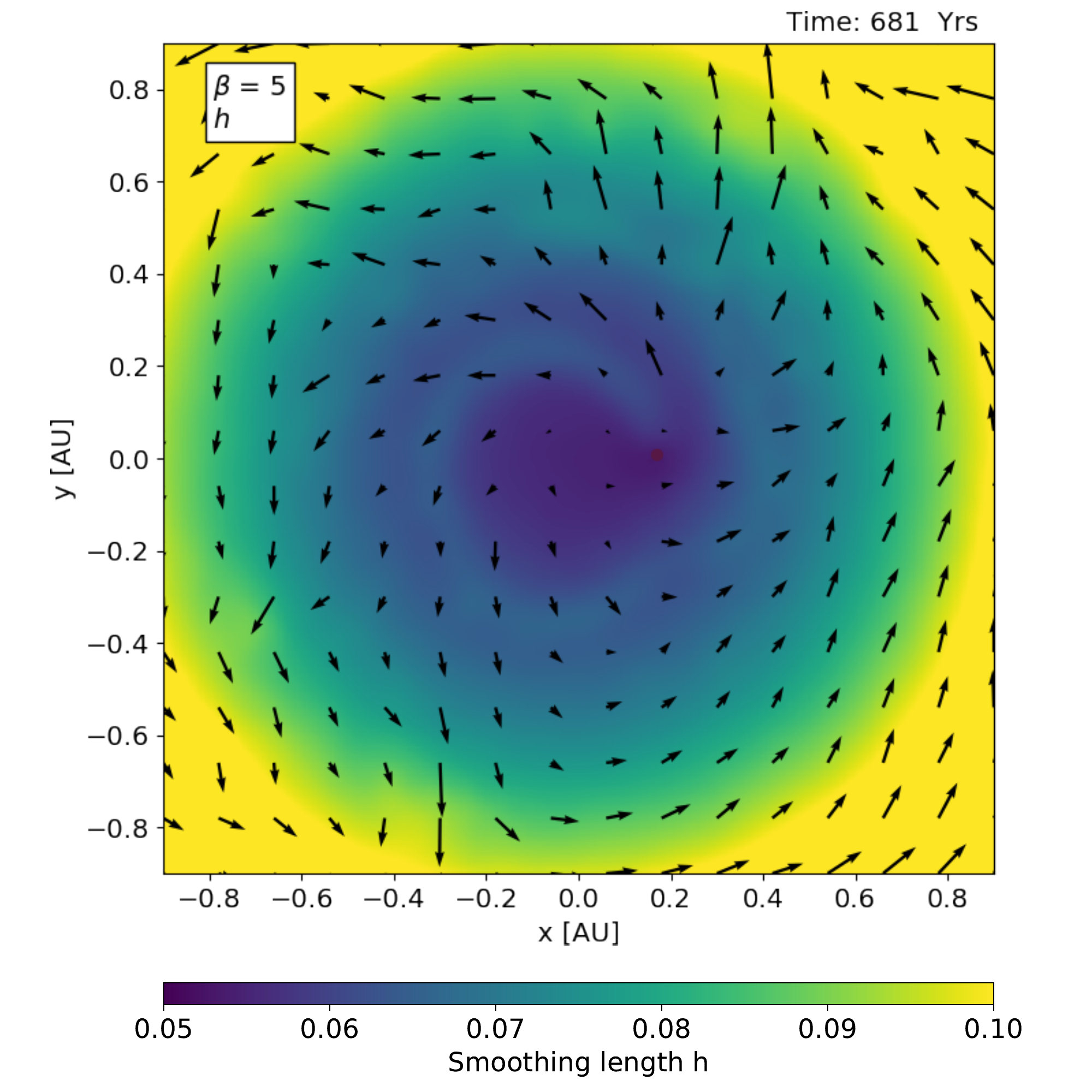}
\caption{A visualisation of the gas smoothing length inside the protoplanet from Section \ref{sec:results}. There is about a 10\% difference in the smoothing length inside and outside the hot trail. The trail width is approximately set by the smoothing length, indicating that the core motion is dependent on our numerical resolution.}
\label{fig:zoom_h}
\end{figure}

In Section \ref{sec:core_modelling} we showed that our cores were accelerated away from the protoplanet centre by a low density, hot gas tail. We also showed that the magnitude of this acceleration was limited by the feedback injection lengthscale in our simulations. In order to demonstrate this conclusively, in this Appendix section we vary the injection lengthscale and study the effect on the core velocity. 

Figure \ref{fig:fb_NN} is similar to Figure \ref{fig:fb_vel}, but now we vary the number of nearest neighbours ($NN_{FB}$) that the feedback is passed to. The fiducial value in the paper was 160. Due to the three dimensional nature of the problem, the feedback injection lengthscale ($l_{FB}$) varies as $l_{FB} \propto NN_{FB}^{1/3}$. We see from the blue and yellow lines that when $l_{FB}$ is shorter, the core velocity is greater and the core reaches a larger offset inside the protoplanet. This demonstrates qualitatively that a larger injection lengthscale leads to a weaker thermal acceleration, as expected from \cite{MassetVelascoRomero17}.

Figure \ref{fig:zoom_h} shows the spatial SPH smoothing length inside our low resoltion $t_{acc}$ = $10^3$ years protoplanet. We see that the width of the hot gas trail is comparable to the smoothing length of the gas inside it. This indicates that further study of the internal temperature structures of these protoplanets is necessary. Results from the AGN feedback community suggest that this will be a highly costly numerical investigation. 

Regardless of the feedback lengthscale, we find that the ultimate protoplanet disruption time remains very similar, only deviating by $\sim$ 5 \% between runs. Note that the $NN_{FB}$=40 run did not disrupt, the large offset allows dust to collect at the centre of the protoplanet and makes the simulation timestep very short. This is the same behaviour observed for our high resolution run in Figure \ref{fig:zoom_dust}.

The disruption of the protoplanet is governed by the total energy release. The similarity in final disruption time indicates that the core offset does not have a significant impact on the magnitude of energy release.
Since hot gas escapes convectively from close to the core, a small thermal conductivity does not prevent the core from heating the entire protoplanet.

We stress that future work should not neglect the thermal acceleration of these luminous cores. The departure of the core away from the protoplanet centre allows for subsequent accreted dust to collect at the centre of the protoplanet and potentially form secondary cores and debris.

\section{Turning off feedback}
\begin{figure}
\includegraphics[width=0.99\columnwidth]{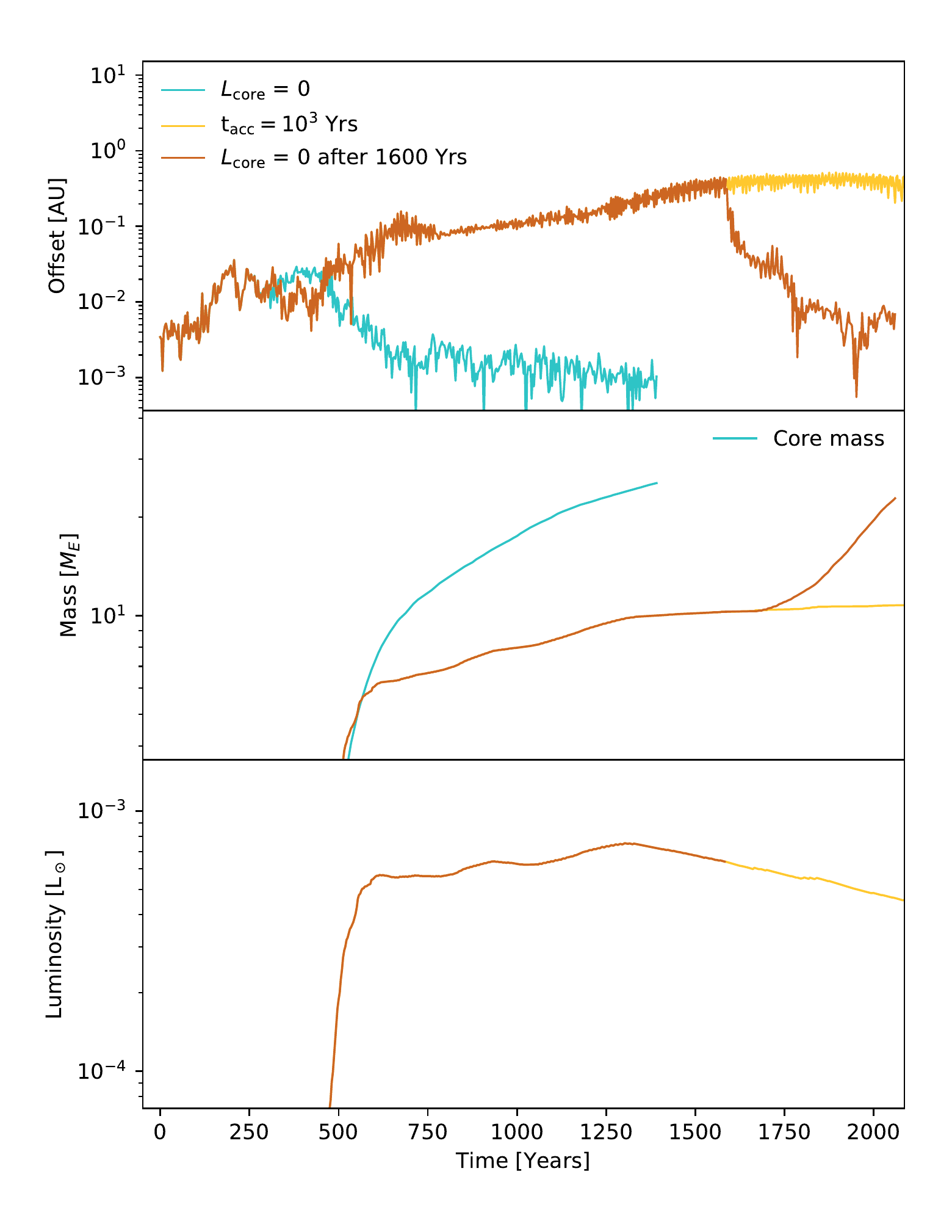}
\caption{Experiments in which we switch of luminosity to demonstrate that the core offset is caused by the heating force. Blue: test case with no feedback, yellow: $t_{\rm acc}=10^3$ years feedback case and orange: feedback `turned off' after 1600 years. After the turn off, the core returns to the centre of the protoplanet and can once again accrete metals.}
\label{fig:fb_shutdown}
\end{figure}

The wandering motion of our cores is very unusual. In order to reassure ourselves that this was linked to the luminous feedback, we ran a simulation in which we `turn off' the feedback part way into the run. The results are shown in Figure \ref{fig:fb_shutdown}. The blue line shows a control case for accretion with no feedback, the core is able to grow rapidly since it remains near the centre of the protoplanet where dust is collecting. For reference, the resolution limit in these simulations is a few $10^{-2}$ AU.  The yellow line plots the standard $t_{\rm acc}=10^3$ years feedback, as seen already the core rapidly leaves the centre of the protoplanet, limiting its subsequent accretion of metals after just 500 years. Finally, the orange line shows a case in which we switch off the feedback after 1600 years.
Once feedback is turned off, dynamical friction on the core causes it to return to it's initial position in the centre of the protoplanet where it can once again start to accrete metals.
These results show that the motion of the core is inextricably linked to our prescription for luminous feedback. Note that this data comes from an earlier simulation setup and so the values do not precisely correspond to those in the main paper.

\section{Radiative diffusion approximation}
We can however make a physical prediction about the importance of this effect on the core by examining how thermal conductivity actually operates inside an optically thick protoplanet.

We can recast the thermal conductivity as the thermal diffusivity ($k$) using the equation $\chi = k /(\rho c_p)$ where $c_p = k_B /(\mu m_P (\gamma-1))$ is the specific heat capacity of the gas. Since we expect the optical depth to be very high in the central regions of protoplanets, we can model the thermal diffusivity with the radiative diffusion approximation

\begin{equation}
k = \dfrac{16}{3}\dfrac{ \sigma_{B}T^3}{\rho \kappa}
\end{equation}

where $\sigma_B$ is the Stefan-Boltzmann constant and $\kappa=\kappa(T,\rho)$ is the opacity\footnote{We use the opacity laws provided in \cite{ZhuEtal09} for this calculation.}. 
In this way we obtain an estimate for the radiative diffusion conductivity in terms of $\rho$, $T$ and $\mu$. Plugging this into Equation \ref{eq:v_MVR17}, we find a very high value for the resulting velocity perturbation, at least two orders of magnitude greater than the sound speed. 

However, at this point the analysis becomes inconsistent since it does not account for the increased local gas temperature due to such a localised injection of energy. If the local gas becomes hotter than $\sim$ 1500 K then it will become optically thin and the velocity perturbation will decrease. Although this is just an estimate, it demonstrates the importance of considering the dynamic effects of core feedback in future research. This is a new paradigm in GI core formation, in previous research GI cores have always been assumed to be fixed at the centre of the gravitational potential well of their parent protoplanet.

\section{Protoplanet internal velocity profiles}
\begin{figure}
\includegraphics[width=0.99\columnwidth]{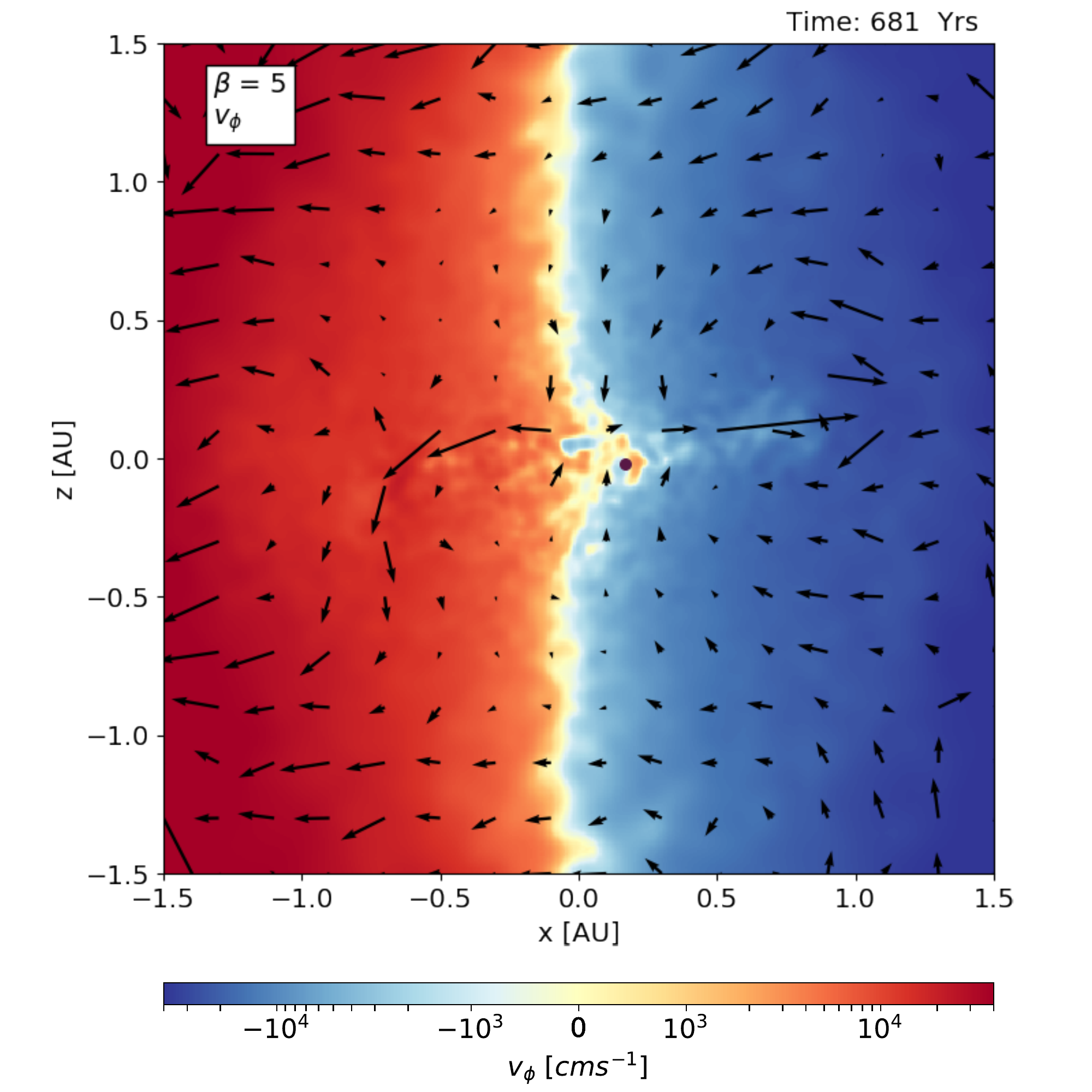}
\caption{Azimuthal velocity field inside the $t_{\rm{acc}}$=1000 years protoplanet, the black dot marks the location of the core. The black arrows plot the radial gas motion, note the outflow of hot gass through the midplane of the protoplanet.}
\label{fig:zoom_vphi}
\end{figure}

\begin{figure}
\includegraphics[width=0.99\columnwidth]{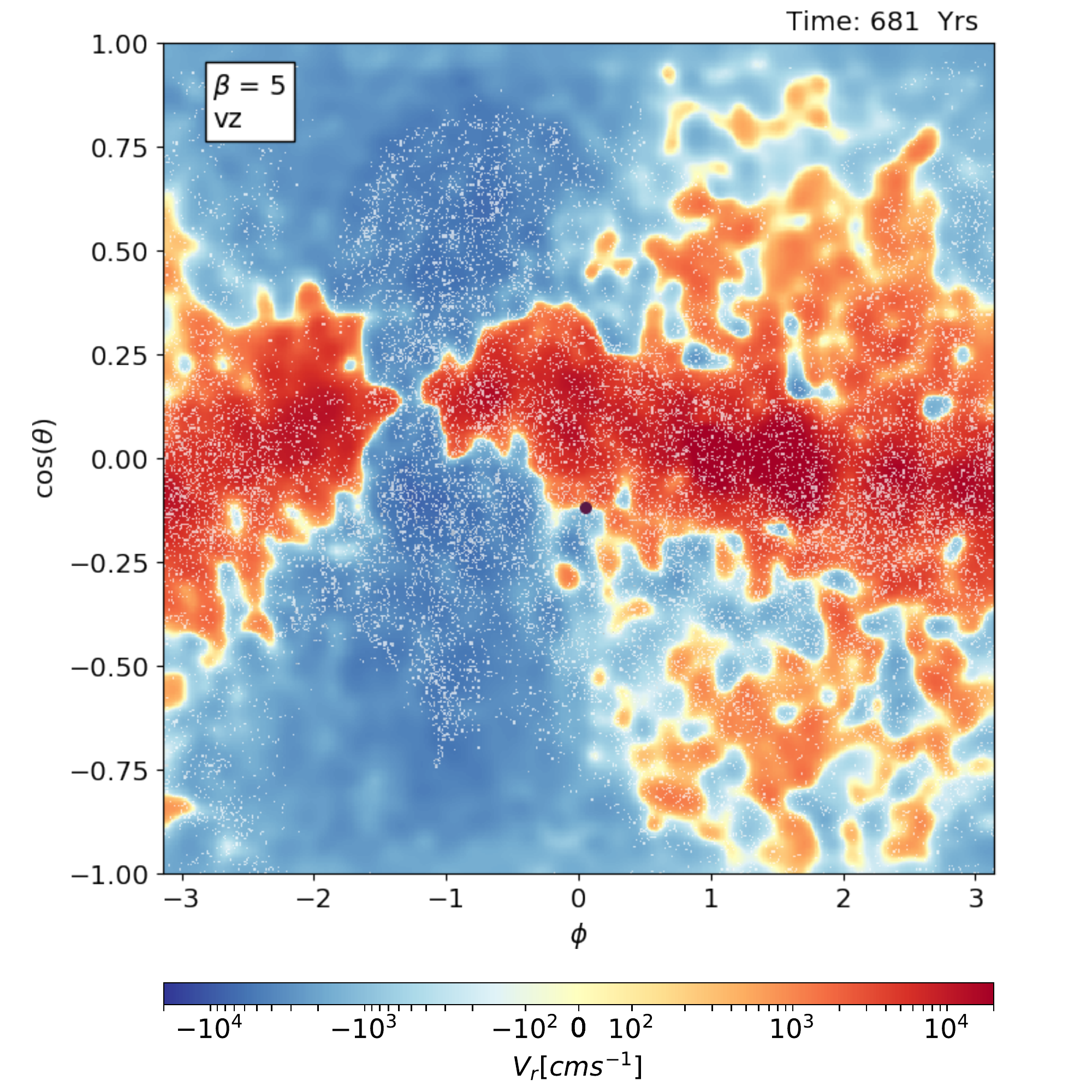}
\caption{A plot of the radial velocity profile at 0.5 AU inside the $t_{\rm{acc}}$=1000 years protoplanet in $\phi$-cos($\theta$) space. Dust is over-plotted as white points and the location of the planet is marked with a blue dot. Note the large contrast between mid-plane outflows driven by core feedback and inflowing gas in the polar regions.}
\label{fig:zoom_vr}
\end{figure}

In the following Appendix section we present some additional figures from the $t_{acc}$ = 1000 years simulation to show the velocity profile of gas inside our protoplanets. They show the system at 681 years, the same time as figures \ref{fig:zoom_rho_xy}-\ref{fig:zoom_T}.

Figure \ref{fig:zoom_vphi} shows the azimuthal velocity profile inside this protoplanet, there is some differential rotation inside the structure but the rotational velocities are much lower than the sound speed which is $\sim 10^5 $cms$^{-1}$.

Figure \ref{fig:zoom_vr} shows the radial velocity profile in $\phi$-cos($\theta$) space at 0.5 AU from the protoplanet centre at the same time as Figures \ref{fig:zoom_rho_xy}-\ref{fig:zoom_T}. The location of the core is marked by a black dot in the centre of the plot. It is off-set from the centre by $R=0.18$ AU, and its nearly circular orbit corresponds to motion from right to left on this figure. We see that the hot gas trail in the x-y plane causes an outflow of gas along the protoplanet midplane which is matched by an inflow of colder gas through the poles. This convective process allows the core to heat the entire protoplanet.
Dust in the protoplanet is over-plotted as white dots, it is broadly concentrated in the x-y plane but the core feedback has caused it to be stirred up in the vertical cos($\theta$) direction.

\bsp	
\label{lastpage}
\end{document}